\documentclass[prd,eqsecnum,twocolumn,amsfonts,showpacs]{revtex4}

\usepackage{graphicx}

\usepackage{bm}

\def\fsl#1{\setbox0=\hbox{$#1$}           
   \dimen0=\wd0                                 
   \setbox1=\hbox{/} \dimen1=\wd1               
   \ifdim\dimen0>\dimen1                        
      \rlap{\hbox to \dimen0{\hfil/\hfil}}      
      #1                                        
   \else                                        
      \rlap{\hbox to \dimen1{\hfil$#1$\hfil}}   
      /                                         
   \fi}                                         %

\newcommand{\beq}{\begin{equation}}
\newcommand{\eeq}{\end{equation}}
\newcommand{\beqs}{\begin{eqnarray}}
\newcommand{\eeqs}{\end{eqnarray}}
\newcommand{\lsim}{\mathrel{\raisebox{-
.6ex}{$\stackrel{\textstyle<}{\sim}$}}}
\newcommand{\gsim}{\mathrel{\raisebox{-
.6ex}{$\stackrel{\textstyle>}{\sim}$}}}

\newcommand{\pslash}{p\hspace{-0.067in}\slash}

\begin{document}

\title{$Z$ Boson Propagator Correction in 
Technicolor Theories with ETC Effects Included} 

\author{Masafumi Kurachi$^{(a)}$}

\author{Robert Shrock$^{(a)}$}

\author{Koichi Yamawaki$^{(b)}$}

\affiliation{
(a) \ C.N. Yang Institute for Theoretical Physics \\
State University of New York \\
Stony Brook, NY 11794}

\affiliation{
(b) \ Department of Physics \\
Nagoya University \\
Nagoya 464-8602, Japan }

\begin{abstract}

We calculate the $Z$ boson propagator correction, as described by the $S$
parameter, in technicolor theories with extended technicolor interactions
included.  Our method is to solve the Bethe-Salpeter equation for the requisite
current-current correlation functions. Our results suggest that the inclusion
of extended technicolor interactions has a relatively small effect on $S$.

\end{abstract}



\maketitle

\vspace{16mm}

\newpage
\pagestyle{plain}
\pagenumbering{arabic}

\section{Introduction}
\label{intro}

The origin of electroweak symmetry breaking (EWSB) is an outstanding unsolved
question in particle physics.  An interesting possibility is that electroweak
symmetry breaking is driven by an asymptotically free, vectorial non-abelian
gauge interaction, technicolor (TC) \cite{tc}, with a coupling that becomes
strong at the electroweak scale.  The EWSB is produced by the formation of
bilinear condensates of technifermions.  To communicate this symmetry breaking
to the standard-model fermions (which are technisinglet), one embeds
technicolor in a larger, extended technicolor (ETC) theory \cite{etc}.  In
order to account for the generational structure of the standard-model (SM)
fermion masses, the ETC gauge symmetry is envisioned to break sequentially in
stages at the respective mass scales $\Lambda_j$, $j=1,2,3$, corresponding to
these generations, finally yielding the technicolor gauge symmetry as an exact,
unbroken subgroup.  Because these scales enter as inverse powers in the
resultant expressions for quark and lepton masses, the highest ETC
symmetry-breaking scale, $\Lambda_1$, corresponds to the first generation, and
so forth for the others.  The scales $\Lambda_j$ and the corresponding masses
of the ETC gauge bosons must be large in order to satisfy constraints from
flavor-changing neutral current processes.  In current, reasonably
ultraviolet-complete, ETC models, these ETC breaking scales are $\Lambda_1 \sim
10^3$ TeV, $\Lambda_2 \sim 50-100$ TeV, and $\Lambda_3 \sim$ few TeV
\cite{at94}-\cite{kt}.  Modern technicolor theories feature a large but slowly
running (``walking'') technicolor gauge coupling, $g_{_{TC}}$
\cite{wtc1}-\cite{chipt3}.  While some early studies of walking assumed an
ultraviolet (UV) fixed point in models with U(1) gauge symmetry, modern walking
technicolor theories are based on the fact that walking can result naturally
from the presence of an approximate infrared (IR) fixed point in the TC
renormalization group equations of the non-abelian technicolor gauge
theory. This occurs at a value $\alpha=\alpha_*$ (where $\alpha =
g_{_{TC}}^2/(4\pi)$) which is close to, but slightly larger than, the minimal
value $\alpha_{cr}$ for which the technifermion condensates form.  An important
property of a technicolor theory with walking behavior is that the anomalous
dimension of the bilinear technifermion operator is $\gamma_{\bar\psi\psi}
\simeq 1$, so that the momentum-dependent dynamical technifermion mass
$\Sigma(p)$ falls off as $p^{-1}$ rather than $p^{-2}$ for an extended range of
Euclidean momenta $p$.  This produces the requisite strong enhancement of SM
fermion masses, relative to the values that they would have in a QCD-like
(non-walking) theory, which is needed in order to fit experiment.  In the pure
technicolor theory, the chiral symmetry breaking occurs if $\alpha C_2(R)$
exceeds a number of order unity, where $C_2(R)$ denotes the quadratic Casimir
invariant for the technifermions, which transform according to the
representation $R$ of the TC gauge group.

Technicolor theories are severely constrained by the corrections that they
induce in precision electroweak quantities, in particular, corrections to the
$Z$ and $W$ boson propagators, conveniently represented by the $S$, $T=\Delta
\rho/\alpha_{em}(m_Z)$, and $U$ \cite{pt} or equivalent \cite{ab} parameters,
where $\rho = m_W^2/(m_Z^2 \cos^2\theta_W)$ and $\Delta \rho$ is the deviation
of $\rho$ from unity due to new physics beyond the standard model.
Experimentally allowed regions in these parameters are given in Refs.
\cite{pdg}.  Since the ${\rm SU}(2)_L \times {\rm U}(1)_Y$ gauge couplings are
small at the TeV scale, the condensates of $T_3=1/2$ and $T_3=-1/2$
technifermions can naturally be approximately equal, so that technicolor
contributions to $\Delta \rho$ are small (e.g., \cite{ssvz}). Further
contributions to $\Delta \rho$ may arise from ETC effects, as discussed below.
From the point of view of low-energy effective field theory, since the
technicolor sector arises as the low-energy residue of the ETC theory, a
reasonable first approximation for calculating $S$ is to consider the
technicolor theory by itself, without any higher-dimension operators arising
from ETC interactions.  Calculations of this type have been performed in
QCD-like and walking technicolor theories \cite{pt}, \cite{scalc}-\cite{sg}
(see also \cite{adscft,csaki}). In particular, it has been found that
technicolor contributions to $S$ may be suppressed in a TC theory with walking
behavior \cite{at_s}-\cite{adscft}.

The question then arises as to what influence the exchange of strongly coupled
massive ETC gauge bosons, and the resultant effective local four-fermion
current-current interaction in the technicolor theory, have on the technicolor
correction to the $Z$ boson propagator, as described by the $S$ parameter.
Since the ETC gauge coupling is strong at the TeV mass scale, the exchange of
the lightest massive ETC vector bosons generates four-fermion operators that
could, {\it a priori}, have a significant effect on the chiral symmetry
breaking in the technicolor theory and on the approximate IR fixed point.  

Accordingly, in this paper we study the effects of ETC-induced four-fermion
operators on $S$.  As a first step, we map out the chiral phase boundary via
the solution of a Schwinger-Dyson equation for the technifermion propagator,
including both massless technigluon exchange and the leading massive ETC gauge
boson exchange.  In the context of a walking technicolor model we then solve a
Bethe-Salpeter equation for the requisite derivative of the current-current
correlation function that yields $S$ (cf. eq. (\ref{scor}) below).  

The effects of various types of four-fermion operators on dynamical chiral
symmetry breaking have been studied in many contexts in the past.  In a
pioneering work, Nambu and Jona-Lasinio (NJL) showed that a four-fermion
operator with a sufficiently strong coupling can induce dynamical chiral
symmetry breaking \cite{njl}.  An early approach to nonperturbative generation
of fermion masses starting with a massless fermion in electrodynamics (QED) was
Ref. \cite{jbw}.  Using large-$N$ methods, Ref. \cite{gn} showed that a certain
four-fermion operator in a $(1+1)$-dimensional model produced dynamical chiral
symmetry breaking. In quantum chromodynamics (QCD) with $N_f=2$ massless quark
flavors, the effective instanton-induced operator is a four-fermion operator
\cite{hooft}, and this was shown to contribute importantly to the formation of
a bilinear quark condensate and associated dynamical chiral symmetry breaking
\cite{qcd_instanton}.  There have also been studies of the chiral phase
transition in both abelian and non-abelian gauge theories with various types of
four-fermion operators \cite{qed4f}-\cite{aoki99}.  We note that the
four-fermion operators that we consider are ETC-induced and differ from
four-fermion operators directly involving top quark condensates, such as appear
in topcolor models.

The paper is organized as follows. Section II is devoted to a review of some
pertinent material on technicolor and extended technicolor theories.  In
Section III we discuss the calculation of the dynamical technifermion mass via
the solution of the Schwinger-Dyson equation and the mapping of the chiral
phase boundary.  Section IV contains the equations expressing $S$ in terms of
current-current correlation functions.  In Section V we discuss our solution of
the Bethe-Salpeter equation and our results for $S$.  Section VI contains our
conclusions.  In an appendix we comment on the similarities and differences
between the ETC-induced four-fermion interaction that we study and the
four-fermion interaction used in the Nambu-Jona Lasino model and its gauged
extensions.

\section{Some Properties of Technicolor and Extended Technicolor Theories}
\label{tcetc}

In this section we discuss some properties of the technicolor and extended
technicolor theories that will be used in our present study. We begin with the
pure technicolor sector and then include ETC.  The technicolor theory is a
vectorial, asymptotically free gauge theory with a gauge group that we take to
have gauge group SU($N_{TC}$), with gauge coupling $g_{_{TC}}$.  We choose
$N_{TC}=2$, as in recent TC/ETC model-building \cite{at94}-\cite{kt}, for
several reasons, including the fact that this choice (i) minimizes technicolor
contributions to the $S$ parameter as compared with larger values of
$N_{_{TC}}$, (ii) can naturally produce a walking technicolor theory in a
one-family model, and (iii) makes possible a mechanism to explain light
neutrino masses \cite{nt}.  To discuss the features of this theory, we shall
revert to general $N_{TC}$ for some formulas.  The theory contains $N_f$
massless Dirac technifermions, and we assume that these transform according to
the fundamental representation of SU($N_{TC}$) \cite{hr}.  The renormalization
group (RG) equation for the running technicolor gauge coupling squared,
$\alpha(\mu) \equiv \alpha_{_{TC}}(\mu)$ is
\beq
\beta = \mu \frac{d \alpha(\mu)}{d \mu} = - \frac{\alpha(\mu)^2}{2\pi}
\left ( b_0 + \frac{b_1}{4\pi}\alpha(\mu) + O(\alpha(\mu)^2) \right ) \ ,
\label{eq:RGE_for_alpha}
\eeq
where $\mu$ is the momentum scale, and $b_0$ and $b_1$ are known coefficients.
The two terms listed are scheme-independent.  The next two higher-order terms
have also been calculated but are scheme-dependent; their inclusion does not
significantly affect our results.  Since the technicolor theory is
asymptotically free, $b_0 > 0$.  For sufficiently large $N_f$, $b_1 < 0$, so
that the technicolor beta function has a second zero (approximate infrared
fixed point of the renormalization group) at a certain $\alpha_*$, given, to
this order, by $\alpha_* = -4\pi b_0/b_1$. As the number of technifermions,
$N_f$, increases, $\alpha_*$ decreases. In a walking technicolor theory, one
arranges so that $\alpha_*$ is slightly greater than the critical value,
$\alpha_{cr}$, for the formation of the bilinear technifermion condensate. As
$N_f$ increases toward $N_{f,cr}$, $\alpha_*$ decreases toward $\alpha_{cr}$
\cite{integer}.  In the one-gluon exchange approximation, the Schwinger-Dyson
equation for the inverse propagator of a technifermion transforming according
to the representation $R$ of the technicolor gauge group yields a nonzero
solution for the dynamically generated fermion mass, (which is an order
parameter for spontaneous chiral symmetry breaking) if $\alpha \ge
\alpha_{cr}$, where $\alpha_{cr}$ is given by
\beq
\frac{3 \alpha_{cr} C_2(R)}{\pi} = 1 \ . 
\label{alfcrit}
\eeq
For the case at hand, where the technifermion transforms according to the
fundamental representation of SU($N_{TC}$), this is $C_2(fund.) \equiv
C_{2F}=(N_{TC}^2-1)/(2N_{TC})$, so that, with $N_{TC}=2$, $\alpha_{cr} \simeq
1.4$.  To estimate $N_{f,cr}$, one solves the equation $\alpha_* =
\alpha_{cr}$, yielding the result \cite{chipt2,chipt3} $N_{f,cr} =
2N_{TC}(50N_{TC}^2-33)/[5(5N_{TC}^2-3)]$.  For $N_{TC}=2$ this gives $N_{f,cr}
\simeq 7.9$.  This estimate is clearly rough, in view of the strongly coupled
nature of the physics.  Moreover, the coupling $\alpha_*$ is only an
approximate IR fixed point of the renormalization group, since the
technifermions gain dynamical masses $\Sigma$ and are integrated out in the
effective field theory for energies below $\Sigma$, where the technicolor beta
function consequently has the form for a pure gauge theory.  Effects of
higher-order gluon exchanges have been studied in \cite{alm}.  From earlier
studies in quantum chromodynamics (QCD), it is known that instantons (which are
not directly included in the above-mentioned Schwinger-Dyson equation)
contribute to the formation of a bilinear fermion condensate in a vectorial
gauge theory \cite{qcd_instanton}.  The effect of instantons on the
(zero-temperature) chiral transition in a vectorial gauge theory as a function
of $N_f$ has been studied in Refs. \cite{inst,shuryak}. In principle, lattice
gauge simulations should provide a way to determine $N_{f,cr}$, but the groups
that have studied this have not reached a consensus \cite{lgt}.
Thus, we shall use the value $N_{f,cr} \simeq 8$ for $N_{TC}=2$ but note that
there some uncertainty in the determination of this number, as is expected in
view of the strongly coupled nature of the physics.

We shall focus on technicolor models in which the technifermions comprise one
family with respect to the standard-model gauge group, $G_{SM} = {\rm SU}(3)_c
\times {\rm SU}(2)_L \times {\rm U}(1)_Y$ (e.g., \cite{fs}).  In such a
one-family technicolor model, the technifermions transform as
\beqs
& & Q_L: \ (N_{TC},3,2)_{1/3,L} \cr\cr
& & U_R: \ (N_{TC},3,1)_{4/3,R} \cr\cr
& & D_R: \ (N_{TC},3,1)_{-2/3,R} \cr\cr
& & L_L: \ (N_{TC},1,2)_{-1,L} \cr\cr
& & N_R: \ (N_{TC},1,1)_{0,R} \cr\cr
& & E_R: \ (N_{TC},1,1)_{-2,R} \ , 
\label{1fam_quarks}
\eeqs
where the numbers in the parentheses refer to the representations of ${\rm
SU}(N_{TC}) \times {\rm SU}(3)_c \times {\rm SU}(2)_L$ and the subscripts refer
to weak hypercharge, $Y$. Hence, with $N_w=2$, this type of model contains
$N_f=N_w(N_c+1)=8$ technifermions.  Given the above-mentioned fact that
$N_{TC}=2 \ \Rightarrow \ N_{f,cr} \simeq 8$, it follows that, to within the
accuracy of the two-loop beta function analysis, this technicolor model can
naturally exhibit walking behavior.  Since there are
\beq
N_D=(N_c+1)
\label{nd}
\eeq
SU(2)$_L$ doublets for each technicolor index, the total number of electroweak
doublets, $N_{D,tot.}$, of technifermions is
\beq
N_{D,tot.}=N_D \, {\rm dim}(R) = (N_c+1)N_{TC}=8 \ , 
\label{ndtot}
\eeq
where $R$ denotes the technifermion representation of SU($N_{TC}$), with 
$R=fund.$ here.

We next discuss the embedding of this technicolor theory in extended
technicolor.  Although much of our analysis of chiral symmetry breaking and the
$S$ parameter is rather general, it will be useful to have a specific class of
ETC models in mind as a theoretical framework.  A natural formulation of ETC
gauges the fermion generational index and combines it with the technicolor
gauge index, so that, for an SU($N_{ETC}$) gauge group, one has
\beq
N_{ETC} = N_{gen.} + N_{TC} = 3 + N_{TC} \ . 
\label{netc}
\eeq
With $N_{TC}=2$, this then leads to SU(5) for the ETC gauge group.  
The SM fermions and corresponding technifermions transform according to the
representations
\beqs
Q_L: \ (5,3,2)_{1/3,L} \ , & & \quad \  u_R: \ (5,3,1)_{4/3,R} \ , \cr\cr
& &                            \quad \  d_R: \ (5,3,1)_{-2/3,R} 
\label{qreps}
\eeqs
and
\beq
L_L: \ (5,1,2)_{-1,L}, \quad\quad  e_R: \ (5,1,1)_{-2,R} \ ,
\label{lreps}
\eeq
where the subscripts denote $Y$. For example, writing out the components of
$e_R$, one has $e_R \equiv (e^1, e^2, e^3, e^4, e^5)_R \ \equiv \ (e, \mu,
\tau, E^4, E^5)_R$, where the last two entries are the charged technileptons.
There are also SM-singlet, ETC-nonsinglet fields in various representations of
SU(5)$_{ETC}$ such that the overall ETC theory is a chiral gauge theory.  The
right-handed, SM-singlet, neutrino fields and corresponding technineutrinos
arise as certain components of these SM-singlet, ETC representations, as
discussed in Refs. \cite{at94}-\cite{kt}.  In this type of ETC theory, because
the SM fermions and the technifermions in each of the ETC multiplets
(\ref{qreps}) and (\ref{lreps}) transform in the same way under the SM gauge
group $G_{SM}$, it follows that the ETC gauge bosons are SM-singlets, and
$[G_{SM},G_{ETC}]=0$.  It may be noted that one could also consider a
technicolor theory with a single electroweak doublet of technifermions, so that
$N_{D,tot.}=N_{TC}$.  Although this model, by itself, does not exhibit walking
behavior, this can be achieved by adding the requisite number of SM-singlet
technifermions, as, e.g., in \cite{ts}.  To avoid electroweak gauge anomalies,
the technifermion electroweak doublet must have weak hypercharge $Y=0$, and
consequently, the ETC gauge bosons that transform SM fermions to these
technifermions carry hypercharge and charge (and, for some, also color), so
that $[G_{SM},G_{ETC}] \ne 0$.  Consequently, an analysis of electroweak
corrections in this type of theory is more complicated than the analysis in a
model based on one-family technicolor, and we do not pursue this here.

The overall ETC theory is constructed to be asymptotically free, so that
as the energy scale decreases from large values, the ETC gauge coupling grows,
and produces various bilinear condensates.  Since the ETC theory is chiral,
these condensates generically break the ETC gauge symmetry, and this can be
arranged to occur in stages.  Explicit, reasonably ultraviolet-complete ETC
models of this type were studied in Refs. \cite{at94}-\cite{kt}. The ETC
gauge bosons may be denoted $V^i_j$, where $1 \le i,j \le 5$.  It is convenient
to use the notation $V_{dj}$, $j=1,2,3$ for the ETC gauge bosons corresponding
to the diagonal Cartan generators $T_{dj}$.  For the sake of generality, we
display the $T_{dj}$ for arbitrary $N_{TC}$.  With the canonical normalization 
${\rm Tr}(T_iT_j)=(1/2)\delta_{ij}$ and in our basis with ETC indices 
ordered as $i=1,2,3$ for generations and $i=\tau = 4,5,...,N_{ETC}$ for 
technicolor, these Cartan generators are 
\beq
T_{d1} = \nu_1 \, {\rm diag}(-(N_{TC}+2),1,1,\{1\}) \ , 
\label{td1}
\eeq
\beq
T_{d2} = \nu_2 \, {\rm diag}(0,-(N_{TC}+1),1,\{1\}) \ , 
\label{td2}
\eeq
and
\beq
T_{d3} = \nu_3 \, {\rm diag}(0,0,-N_{TC},\{1\}) \ , 
\label{td3} 
\eeq
where $\{1\}$ denotes a string of $N_{TC}$ 1's, 
\beq
\nu_1 = [2(N_{TC}+2)(N_{TC}+3)]^{-1/2} \ , 
\label{nu1}
\eeq
\beq
\nu_2 = [2(N_{TC}+1)(N_{TC}+2)]^{-1/2} \ , 
\label{nu2}
\eeq
and, most importantly for our purposes, 
\beq
\nu_3 = [2N_{TC}(N_{TC}+1)]^{-1/2} \ . 
\label{nu3}
\eeq
In particular, $\nu_3=1/(2\sqrt{3})$ for $N_{TC}=2$. 

At the scale $\Lambda_1 \simeq 10^3$ TeV, the SU(5)$_{ETC}$ gauge symmetry is
envisioned to break to SU(4)$_{ETC}$, with the nine ETC gauge bosons in the
coset SU(5)$_{ETC}$/SU(4)$_{ETC}$, $V^1_j$, $V^j_1=(V^1_j)^\dagger$, $2 \le j
\le 5$, and $V_{d1}$, picking up masses $M_1  \simeq \Lambda_1$.  Similarly, at
the scale $\Lambda_2 \simeq 50-100$ TeV, SU(4)$_{ETC}$ breaks to SU(3)$_{ETC}$,
with the seven ETC gauge bosons in the coset SU(4)$_{ETC}$/SU(3)$_{ETC}$,
$V^2_j$, $V^j_2=(V^2_j)^\dagger$, $3 \le j \le 5$, and $V_{d2}$, picking up
masses $M_2 \simeq \Lambda_2$.  Finally, at the scale $\Lambda_3 \simeq $ few
TeV, SU(3)$_{ETC}$ breaks to the residual exact technicolor gauge group
SU(2)$_{TC}$, and the five ETC gauge bosons in the coset
SU(3)$_{ETC}$/SU(2)$_{TC}$, $V^3_j$, $V^j_3=(V^2_j)^\dagger$, $j=4,5$, and
$V_{d3}$ gain masses $M_3 \simeq \Lambda_3$.  Henceforth, we shall denote
technicolor indices as $\tau=4,5$ to distinguish them from generational indices
$i=1,2,3$.  In principle, other strongly coupled gauge symmetries such as
topcolor might also be present, but we will not consider these here.

The most important ETC contributions to the technicolor sector arise from the
exchange of the lowest-lying massive ETC gauge bosons, namely those with mass
$M_3$.  Exchanges of more massive ETC gauge bosons make contributions that are
strongly suppressed by factors $M_3^2/M_j^2 \ll 1$, where $j=1,2$.
In the technicolor theory, as a low-energy effective field theory, the exchange
of massive ETC gauge bosons produce local four-fermion operators of the
current-current form.  There are two types of corrections to the propagator of
a technifermion due to the emission and reabsorption of virtual ETC gauge
bosons of mass $M_3$, namely those involving (i) $V_{d3}$ and (ii) $V^\tau_3$.
Thus a one-loop technifermion propagator correction involving the exchange (i)
has the same technifermion on the internal fermion line, while that involving
the exchange (ii) has the corresponding third-generation SM fermion on the
internal line.  In a viable ETC model, the exchange (ii) must make a smaller
contribution to the dynamical mass $\Sigma_F$ of a technifermion $F$ 
than the exchange (i) because it violates custodial symmetry.  This is a
consequence of the fact that the emission and reabsorption of a virtual
$V^\tau_3$ yields a one-loop diagram in which a $U$ techniquark transforms to a
virtual $t$ quark and back, while a $D$ techniquark transforms to a $b$ quark
and back.  This correction to the dynamical mass $\Sigma$ of the techniquark
therefore introduces a dependence on $m_t$ for $U$ and $m_b$ for $D$, so
$\Sigma_U$ would, in general, differ significantly from $\Sigma_D$, violating
custodial symmetry in the technicolor sector.  

  Such violations must be small.  Global fits to data yield allowed regions in
$(S,T)$ depending on a reference value of the SM Higgs mass, $m_{H,ref.}$. The
comparison of these with a technicolor theory is complicated by the fact that
technicolor has no fundamental Higgs field. Sometimes one formally uses
$m_{H,ref.} \sim 1$ TeV for a rough estimate, since the standard model with
$m_H \sim 1$ TeV has strong longitudinal vector boson scattering, as does
technicolor. However, this may involve some double-counting when one also
includes contributions to $S$ from technifermions, whose interactions and bound
states (e.g., techni-vector mesons) are responsible for the strong scattering
in the $W^+_L W^-_L$ and other longitudinal vector-vector channels in a
technicolor framework.  The current allowed region in $(S,T)$ disfavors values
of $S \gsim 0.1$ and $T \gsim 0.4$ \cite{pdg}.  

Because violations of custodial symmetry in the technicolor sector must be
small for the theory to be viable, we shall focus on the corrections of type
(i).  This constraint also implies that $|\Sigma_U - \Sigma_D|/(\Sigma_U +
\Sigma_D) \ll 1$, so we shall drop the subscripts on $\Sigma_U$ and $\Sigma_D$.
Indeed, since all SM interactions are small at the scale of technicolor mass
generation, it is expected that the dynamical masses for all of the
technifermions, $\Sigma_U$, $\Sigma_D$, $\Sigma_N$, and $\Sigma_E$, are
approximately equal, and we shall therefore simply denote them as $\Sigma$.

In the ETC framework that we use, the $V_{d3}$ exchange yields an effective
local operator in the technicolor theory of the form
\beqs
& & {\cal L}_{V_{d3}} = -\frac{\nu_3^2 \, g_{_{ETC}}^2}{M_3^2} \, \sum_\psi
\Big [\sum_{\tau} \bar\psi_\tau    \gamma_\mu \psi^\tau \Big ] 
\Big [\sum_{\tau'}\bar\psi_{\tau'} \gamma^\mu \psi^{\tau'}\Big ] \ , \cr\cr
& & 
\label{etcjj} 
\eeqs
where $\sum_\psi$ is over the technifermions $\psi$ in the theory, and
$\sum_\tau$ is over the technicolor indices.  As is evident from
eqs. (\ref{nu3}) and (\ref{etcjj}), the dimensionless coefficient $\nu_3
g_{_{ETC}}^2$ is not independent of the technicolor theory and its coupling,
$g_{_{TC}}$, since these arise from the sequential breaking of the ETC theory.
However, from an abstract field-theoretic point of view, it is of interest to
investigate the (zero-temperature) chiral transition of an asymptotically free
vectorial gauge theory in the presence of a four-fermion operator with a
coupling that can be varied independently of the gauge coupling.  In this
context, one could map out the chiral phase boundary as a function of these
two, {\it a priori} independent, dimensionless couplings.  If, indeed, one
really posited a four-fermion operator as a fundamental interaction in the
theory, it would change the renormalization-group behavior of the couplings.  
However, in the physical ETC context, in which this four-fermion operator is
the low-energy remnant of the ETC theory, its effects on the technicolor 
gauge coupling can be considered to have already been taken into account at the
higher scale where the ETC gauge degrees of freedom are dynamical, since the 
technicolor sector emerges as the low-energy effective field theory from this
larger ETC theory.

By formally varying $M_3$ from zero to its physical nonzero value, and noting
that the sign of the resultant contribution to the fermion propagator does not
change, one infers that the effect of the ETC interaction is to enhance the
spontaneous chiral symmetry breaking.  Indeed, the NJL model showed that a
four-fermion operator can induce dynamical chiral symmetry breaking by itself
if its coupling strength is sufficiently great \cite{njl}.  Note that 
ETC-induced four-fermion operator (\ref{etcjj}) is a product of two vector
currents and is different from the four-fermion operator that appears in the 
gauged NJL model, namely a sum of the form $VV - AA$, where $V$ and $A$ denote
vector and axial vector currents.

\section{Schwinger-Dyson Equation and Mapping of Phase Boundary} 
\label{sd}

The first step in our analysis is the calculation of the dynamical
technifermion mass and the mapping of the associated chiral phase boundary in
the theory.  Some early studies using Schwinger-Dyson equations and related
methods to study nonperturbative fermion mass generation in abelian and
non-abelian gauge theories are Refs. \cite{jbw}, \cite{cjt}.
Past work on the (zero-temperature) chiral transition for abelian and
non-abelian gauge models in the presence of four-fermion operator of various
types is contained in Refs. \cite{kmy}-\cite{holdom4f}.  Several of these
studies used models with a (non-asymptotically free) U(1) gauge symmetry.  In
these studies it was assumed that the theory had a UV fixed point at
$\alpha_{cr}^{(0)}$.  This contrasts with modern TC/ETC models, in which the TC
sector arises as the low-energy limit of an ETC theory, and both are based on
asymptotically free, non-abelian gauge symmetries, with $\alpha_{cr}^{(0)}$
being an approximate IR fixed point of the TC theory.  

  Our method is to use the Schwinger-Dyson (SD) equation for the technifermion
propagator to calculate the dynamically induced mass $\Sigma$, taking into
account the exchange of both massless technigluons and the massive $V_{d3}$ ETC
vector bosons.  The dynamical technifermion mass serves as an order parameter
for the chiral transition.  This mass may be defined in terms of the
momentum-dependent fermion mass evaluated at an appropriate Euclidean reference
momentum.  One choice would be $\Sigma = \Sigma(p_E=0)$, but we shall actually
use the related choice
\beq
\Sigma = \Sigma(p_E=\Sigma) \ . 
\label{sigmadef}
\eeq
We recall that the analysis of the Schwinger-Dyson equation, by itself, does
not give direct information about whether or not the theory has confinement.
Indeed, the original NJL model provides an example of dynamical chiral symmetry
breaking via a four-fermion operator (with sufficiently large coupling of the
correct sign) without confinement.  In the pure gauge theory, for $N_f <
N_{f,cr}$, i.e., for $\alpha_* > \alpha^{(0)}_{cr}$, one can argue persuasively
that (although the area law behavior of the Wilson loop ceases to hold because
of string breaking) the theory exhibits confinement, by a formal analytic
continuation in $N_f$ from $N_f=0$ \cite{integer}. This is all that we will
need for our present purposes.  In the pure gauge theory, the chirally
symmetric phase is expected to be a non-abelian deconfined Coulombic phase
for sufficiently weak coupling.  If appropriate conditions were satisfied,
there might possibly also be an intermediate phase with confinement but no
spontaneous chiral symmetry breaking; the presence or absence of this phase
will not be important for our work here.  Since for physical reasons, viz., 
the necessity of a nonvanishing technifermion condensate for electroweak
symmetry breaking, we must be in the confined phase.

\begin{figure}
  \begin{center}
    \includegraphics[height=1cm,width=8cm]{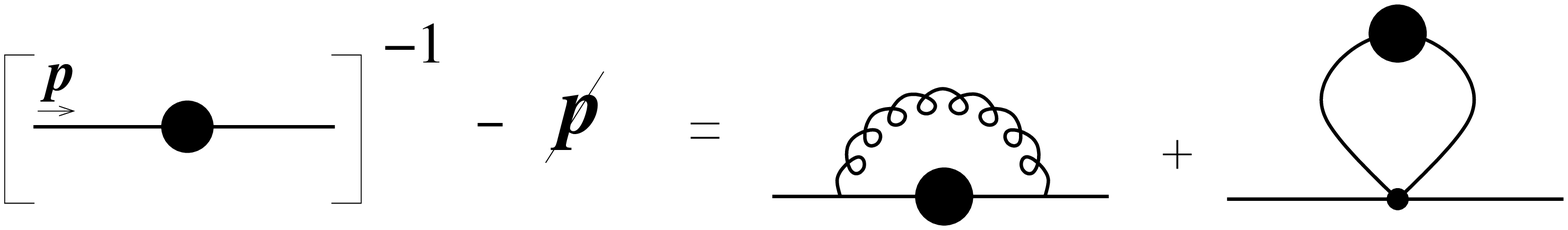}
  \end{center}
\caption{Pictorial representation of the Schwinger-Dyson equation for the
  technifermion.  The first graph on the right represents technigluon exchange,
  and the second represents the contribution of the effective local
  four-fermion operator resulting from the exchange of the massive ETC gauge
  boson $V_{d3}$.}
\label{SDeq}
\end{figure}

The full inverse technifermion propagator can be written as 
\beq
S_F(p)^{-1} = A(p^2) \pslash - B(p^2) \ . 
\label{sf}
\eeq
The resultant SD equation has the form 
\beq
S_f(p)^{-1} - \pslash = I_{TC} + I_{ETC} \ , 
\label{sdgen}
\eeq
where $I_{TC}$ and $I_{ETC}$ are the contributions from technigluons and
$V_{d3}$ ETC vector bosons, respectively.  A graphical representation of these
contributions is shown in Fig. \ref{SDeq}.  The first graph on the right is the
(massless) technigluon exchange diagram, and the second represents the
contribution of the effective local four-fermion operator resulting from the
exchange of the massive ETC gauge boson $V_{d3}$.  The dark blobs on the
technifermion line signify that we use the full technifermion propagator.  
The first term in eq. (\ref{sdgen}) is given by
\beq
I_{TC} = -C_{2F} \int \frac{d^4 q}{i (2\pi)^4} \ 
g_{TC}^2(p,q) D^{TC}_{\mu\nu}(p-q) \ \gamma^\mu S_F(q)\gamma^\nu
\label{itc}
\eeq
where the technigluon propagator is 
\beq
D^{TC}_{\mu\nu}(k) = \frac{N_{\mu\nu}}{k^2}
\label{delta}
\eeq
with numerator (depending on the gauge parameter $\xi$)
\beq
N_{\mu\nu} = - g_{\mu\nu} + \xi \frac{k_\mu k_\nu}{k^2} \ . 
\label{nmunu}
\eeq

Our analysis is thus an improved ladder approximation, where the term
``improved'' refers to the fact that in eq. (\ref{itc}) we take account of 
the momentum dependence of the running technicolor coupling, $g_{TC}$.
We next explain several further approximations that will be made.  In the pure
technicolor theory, if $N_f$ were greater than $N_{f,cr}$, i.e., if $\alpha_*$
were less than $\alpha_{cr}$, so that this IR fixed point of the two-loop RG
equation were exact, then one could make use of an exact solution to this
equation in the entire energy region.  With $b \equiv b_0/(2\pi)$, this
solution is given by \cite{gardi,exactsolution}
\beq
\alpha(\mu)=\alpha_\ast \left[\ W(e^{-1}(\mu/\Lambda)^{b \alpha_*})+1 \
\right]^{-1},
\label{lambert}
\eeq
where $W(x) = F^{-1}(x)$, with $F(x) = x e^x$, is the Lambert $W$
function, and $\Lambda$ is a RG-invariant
scale defined by~\cite{chipt2}
\beq
 \Lambda \ \equiv\  \mu \ \exp \left[ -\frac{1}{b} \left \{
\frac{1}{\alpha_\ast}
\ln\left( \frac{\alpha_\ast - \alpha(\mu)}{\alpha(\mu)} \right)
        + \frac{1}{\alpha(\mu)} \right \} \right] .
\label{lambda}
\eeq
However, since physically, one must be in the phase where $\alpha >
\alpha_{cr}$ so that the technifermion condensate forms and the electroweak
symmetry is broken, the IR fixed point is only approximate rather than exact,
as noted above.  Hence, eq. (\ref{lambert}) is only applicable in an
approximate manner to our case; for momenta much less than the dynamical
fermion mass $\Sigma$, the fermions decouple, and in this very low-momentum
region, with the fermions integrated out, the resultant $\alpha$ would increase
above the value $\alpha_*$ at the approximate IR fixed point.  But since
$\Sigma \ll \Lambda$ in a walking theory, it follows that this lowest range of
momenta makes a relatively small contribution to the integrals to be evaluated
in our calculations.  Hence, over most of the integration range for these
integrals where the coupling $\alpha$ is large, it is approximately constant
and equal to its fixed-point value, $\alpha_*$.  This means that one can use
the approximation
\beq
  \alpha(\mu) = \alpha_\ast \, \theta(\Lambda-\mu) \ ,
\label{stepfunction}
\eeq
where $\theta$ is the step function.  Moreover, we shall assume that 
\beq
g_{_{TC}}(p,q) = g_{_{TC}} \Big ( (p-q)^2 \Big ) = g_{_{TC}}(p_E^2+q_E^2) \ . 
\label{gform}
\eeq
Since $g_{_{TC}}$ would naturally depend on the technigluon momentum squared,
$(p-q)^2 = p^2+q^2-2p \cdot q$, the functional form (\ref{gform}) amounts to
dropping the scalar product term, $-2p \cdot q$.  This is a particularly
reasonable approximation in the case of a walking gauge theory because most of
the contribution to the integral (\ref{itc}) comes from a region of
Euclidean momenta where $\alpha$ is nearly constant.  Hence, the shift upward
or downward due to the $-2p \cdot q$ term in the argument of $\alpha$ has very
little effect on the value of this coupling for the range of momenta that make
the most important contribution to the integral.  The approximations
(\ref{stepfunction}) and (\ref{gform}) are the same as in our previous work
\cite{hky_s,s,mmw,pms}.

Making a Euclidean rotation and performing the angular integration in $I_{TC}$
then yield two equations, for $A(p_E^2)$ and $B(p_E^2)$.  In Landau gauge,
with gauge parameter $\xi=1$, the solution to the equation for $A(p_E^2)$ is
$A(p_E^2)=1$, so that the dynamical mass of the technifermion,
$\Sigma(p_E^2)=B(p_E^2)/A(p_E^2)$, takes the simple form
$\Sigma(p_E^2)=B(p_E^2)$.  This simplification motivates the use of Landau
gauge, although physical results involving $\Sigma$ are, of course, invariant
under technicolor gauge transformations (e.g., \cite{amnw,alm}).  For the
integral $I_{TC}$, setting $x \equiv p_E^2$ and $y \equiv q_E^2$, we obtain
\beq
I_{TC} = \frac{3 C_{2F}}{16 \pi^2} \int_0^\infty y \, dy \, 
\frac{g_{TC}^2(x+y) \, \Sigma(y)}{\max(x,y) \, [y + \Sigma^2(y)]} \ .
\label{sdeq}
\eeq
Note that although the upper limit on the integration is formally infinite, the
integral is actually cut off at $y \simeq \Lambda^2$ because of
eq. (\ref{stepfunction}).  If the technigluon exchange were the only
contribution in eq. (\ref{sdgen}), then this equation would have a nonzero
solution for $\Sigma(p_E^2)$ if $\alpha_{_{TC}} > \alpha^{(0)}_{cr}$, where
$\alpha^{(0)}_{cr}$ was given by eq.  (\ref{alfcrit}).  In our calculations, we
consider the full nonlinear integral equation (\ref{sdeq}).  However, for
comparison with our numerical results, we recall in the pure technicolor theory
(without the ETC-induced four-fermion interaction), if, as $\alpha_* \searrow
\alpha^{(0)}_{cr}$, one neglects the momentum dependence of $\Sigma(y)$ in the
denominator of eq. (\ref{sdeq}), then the solution is
\cite{chipt2}-\cite{chipt3}
\beq
\Sigma = {\rm const.} \,  \Lambda \, \exp
\bigg [ -\pi \Big ( \frac{\alpha_*}{\alpha_{cr}} - 1 \Big )^{-1/2} \bigg ] \ ,
\label{sigsol}
\eeq
A numerical solution of the Schwinger-Dyson equation in a non-abelian gauge
theory found a rather similar result, $\Sigma \propto \Lambda \exp[-0.82
\pi(\alpha_*/\alpha_{cr}-1)^{-1/2}]$ \cite{chipt3}.  A similar numerical
solution of the Schwinger-Dyson equation for a range of values of $\alpha$
slightly larger than $\alpha^{(0)}_{cr}$ obtained results which were fit with
the form (\ref{sigsol}) \cite{mmw}.

We next discuss the leading ETC contribution to eq. (\ref{sdgen}), $I_{ETC}$.
We again shall introduce some physically motivated approximations to simplify
the calculation.  In models with explicit specification of ETC dynamics (e.g.,
\cite{at94}-\cite{kt}), one finds that the walking regime typically extends
from the TC scale to the lowest ETC symmetry breaking scale,
$\Lambda_3$. Hence, as far as the technicolor theory is concerned, the scale
parameter $\Lambda$ in eq. (\ref{lambda}) is of order $\Lambda_3$.  At momentum
scales $\mu \gsim \Lambda_3$, one is dealing with the full SU(3)$_{ETC}$ gauge
interaction (and so forth on up to SU(5)$_{ETC}$ for $\mu \gsim \Lambda_1$).
Although the dynamical gauge degrees of freedom in the coset ${\rm
SU}(3)_{ETC}/{\rm SU}(2)_{TC}$ are frozen out for momenta $\mu \lsim
\Lambda_3$, it will be convenient to use the Landau-gauge form of the ETC gauge
boson propagator for our calculation, so that
\beq
I_{ETC} = -\nu_3^2 \int \ \frac{d^4 q}{i (2\pi)^4} \
g_{_{ETC}}^2(p,q) D^{ETC}_{\mu\nu}(p-q) \ \gamma^\mu S_F(q)\gamma^\nu
\label{ietc}
\eeq
where 
\beq
D^{ETC}_{\mu\nu}(k) = \frac{N_{\mu\nu}}{k^2 - M_3^2} \ . 
\label{delta_etc}
\eeq
This choice maintains $A(p_E^2)=1$. We can then combine the TC and ETC terms as
$I=I_{TC}+I_{ETC}$, with
\beq
I = -\int \ \frac{d^4 q}{i (2\pi)^4} \, 
\frac{\kappa(p,q)}{(p-q)^2}N_{\mu\nu} \ \gamma^\mu S_F(q)\gamma^\nu
\label{i}
\eeq
where
\beqs
& & \kappa(p,q) = C_{2F} \, g_{TC}^2(p,q) + 
\nu_3^2 \, g_{ETC}^2(p,q) \Big [ \frac{(p-q)^2}{(p-q)^2-M_3^2} \Big ] \ . 
\cr\cr
& & 
\label{kappa}
\eeqs
Since the dominant contribution to the momentum integration in $I$ is from
scales smaller than $M_3$, the momentum-dependent term in the denominator of
the second term (\ref{lambda}) is dropped relative to $M_3^2$.  Since, as
noted, the TC sector is the low-energy effective theory arising from ETC, the
gauge couplings are closely related.  As noted, in explicit ETC models, the
walking regime for the TC theory typically extends up to the scale $\Lambda_3$
where the ETC symmetry breaks to the TC symmetry.  Thus, the approximation
(\ref{stepfunction}) for $g_{_{TC}}(\mu)$ means that the Euclidean momentum
integration in the integrals is cut off at $\simeq \Lambda_3$.  Below this
scale, the gauge degrees of freedom in the coset ${\rm SU}(N_{TC}+1)/{\rm
SU}(N_{TC})$ are frozen out, and $g_{_{ETC}}$ does not run.  Combining this
with the fact that the TC gauge coupling inherits its magnitude from the ETC
gauge coupling, we will approximate $g_{_{ETC}}(\mu)$ by the same form as
$g_{_{TC}}(\mu)$, given in eq. (\ref{stepfunction}), namely $g_{_{ETC}}(\mu)
\simeq r g_{_{TC}}(\mu)$, with the parameter $r \simeq 1$ introduced to account
for a slight difference in magnitude between these couplings.  We thus obtain
(with $x=p_E^2$ and $y=q_E^2$)
\beq
\Sigma(x) = \frac{3}{16\pi^2} \int_0^\infty y \, dy \, \frac{\kappa(x+y) \, 
\Sigma(y)} {{\rm max}(x,y) \, [y+\Sigma(y)^2]} \ . 
\label{sigeq}
\eeq
(Note again that the integral is cut off at $y \simeq \Lambda^2$ because of 
eq. (\ref{stepfunction}).)  From the above, the explicit form for $\kappa(z)$
(where $z=x+y$), is
\beq
\kappa(z) = g_{_{TC}}^2(z) \Big [ C_{2F}+r^2 \, \nu_3^2 \frac{z}{M_3^2} \Big ]
\ . 
\label{kappaz}
\eeq
Including the prefactor $3/(16\pi^2)$, we can write this as 
\beq
\frac{3 \kappa(z)}{16\pi^2} = \frac{\alpha_{_{TC}}(z)}{4\alpha^{(0)}_{cr}}
+ \kappa_4(z) \, \frac{z}{M_3^2} \ , 
\label{fullkappa}
\eeq
where the rescaled coefficient of the ETC-induced four-fermion coupling, 
$\kappa_4(z)$, is 
\beq
\kappa_4(z) = \frac{3 r^2 \nu_3^2 \, \alpha_{_{TC}}(z)}{4\pi} \ . 
\label{lambdabar}
\eeq
With $N_{TC}=2$ and hence $\nu_3^2 = 1/12$, if $\alpha_{_{TC}} \simeq
\alpha^{(0)}_{cr} = \pi/(3C_{2F})$, then, taking $r \simeq 1$, the value of the
momentum-dependent $\kappa_4$ at the relevant scale of order a TeV, which we
shall refer to simply as $\kappa_4$, is
\beq
\kappa_4 \simeq \frac{1}{48C_{2F}} \simeq 0.03 \ . 
\label{lambdabar_value}
\eeq
Thus, the value of $\kappa_4$ in the type of ETC model considered here is
rather small.  

 In addition to ETC-induced four-fermion interactions, certain four-fermion
operators could be induced by the nonperturbative dynamics of the technicolor
theory itself.  In particular, as we discussed above, instanton effects produce
effective local multifermion operators, and, upon contraction of technifermion
fields, these yield a particular type of four-fermion operator, which has been
shown to be important for chiral symmetry breaking in QCD \cite{qcd_instanton}.
These instanton-generated multifermion operators are soft, i.e., the mass that
enters as an inverse square factor multiplying them in an effective Lagrangian
is of order the QCD chiral symmetry-breaking scale.  Indeed, NJL-type
four-fermion interactions are commonly used in modern phenomenological models
of chiral symmetry breaking in QCD at zero and finite-temperature
\cite{qcdnjl}. Instanton effects on the chiral phase transition depending on
$N_f$ have been studied in Refs. \cite{inst,shuryak}.  In
Refs. \cite{ksy,aoki99} it was suggested that a four-fermion interaction with a
strength equivalent to our $\kappa_4 \simeq 1/4$ could be induced by the
non-perturbative dynamics of walking technicolor itself.  As with
instanton-induced multifermion operators, the mass scale characterizing these
four-fermion interactions is expected to be $\simeq \Sigma$.  This is different
from the type of ETC-induced four-fermion operator considered here, which is
hard at the scale $\Sigma$ (and become soft above $\Lambda_3$).  Although there
are no instantons in QED$_4$, the importance of four-fermion operators has also
been discussed in connection with a possible UV fixed point in this theory in
Ref. \cite{qed4f}.  (This question was relevant to the interpretation of
differing results from lattice gauge theory simulations concerning the
continuum limit of 4D \ U(1) lattice gauge theory \cite{kogut,schierholz}.) If
the technicolor theory itself generates large four-fermion operators, then the
properties of the theory, even in the absence of ETC effects, might correspond
to an interval effectively equivalent to our $1/4 \lsim \kappa_4 \le 1$ in the
phase diagram of Fig. 7 below.  Apart from this, there is also interest, from a
general field-theoretic point of view, in investigating the range of $\kappa_4$
values extending up to O(1), where the four-fermion coupling, by itself, would
be sufficient to break the chiral symmetry.  In accordance with our discussion
above, we take $M_3$ to be equal to the scale $\Lambda$ that effectively sets
the upper limit of the walking regime for the technicolor theory.

\begin{figure}
  \begin{center}
    \includegraphics[height=8cm,width=7.5cm]{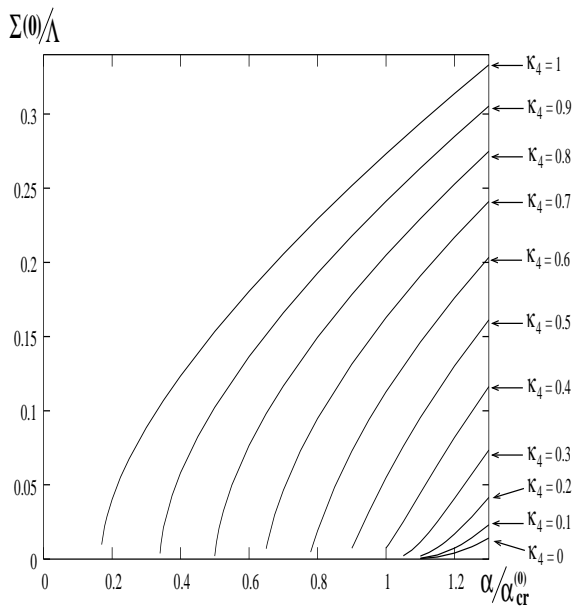}
  \end{center}
\caption{Plot of $\Sigma(0)/\Lambda$ as a function of 
$\alpha/\alpha^{(0)}_{cr}$, for various values of $\kappa_4$.}
\label{Sigma0}
\end{figure}

One next discretizes the Schwinger-Dyson equation and solves it using iterative
numerical methods, as described in Ref. \cite{mmw} (which also contains
references to the literature Schwinger-Dyson and Bethe-Salpeter equations).  In
this analysis, one formally takes $\alpha$ and $\kappa_4$ to be independent,
although, as discussed above, in the actual ETC context, they are related. 
In Fig. \ref{Sigma0} we show the solution for the dynamical technifermion mass
$\Sigma$ (divided by $\Lambda$) as a function of $\alpha/\alpha^{(0)}_{cr}$ for
a range of values of $\kappa_4$.  The effect of the local current-current
interaction is clearly to enhance the spontaneous chiral symmetry breaking, so
that the critical value of the technicolor coupling for the appearance of a
nonzero $\Sigma$, which can be denoted $\alpha_{cr}$, decreases monotonically
as $\kappa_4$ increases from zero.  Similarly, if one fixes the value of
$\alpha/\alpha^{(0)}_{cr}$ where $\Sigma(0)/\Lambda$ is nonzero, then the value
of $\Sigma(0)/\Lambda$ increases monotonically as $\kappa_4$ increases.  For
$\kappa_4=0$, our results are, as expected, in agreement with the exponential
vanishing in eq. (\ref{sigsol}) as $\alpha$ decreases toward its critical
value, $\alpha^{(0)}_{cr}$. As $\kappa_4$ increases, this behavior changes.
Although the detailed critical behavior as one approaches the chiral boundary
is not the focus of our present work, we comment that if one were to make a fit
of the form
\beq
\frac{\Sigma(0)}{\Lambda} \propto \Big ( \frac{\alpha}{\alpha_{cr}} - 1 \Big
  )^{\beta_\Sigma} \quad {\rm as} \ \ \frac{\alpha}{\alpha_{cr}} \to 1^+ \ , 
\label{betadef}
\eeq
where $\beta_\Sigma$ is a critical exponent, analogous to the standard notation
for the critical exponent for the order parameter in statistical mechanics,
then the value of $\beta_\Sigma$ decreases from its value of infinity for
$\kappa_4=0$, corresponding to the essential zero in eq. (\ref{sigsol}),
through the value $\beta_\Sigma \simeq 1$ at $\kappa_4=0.5$, to the value
$\beta_\Sigma \simeq 0.5$ at $\kappa_4=1$.  

\begin{figure}
\begin{center}
\includegraphics[height=8.5cm,width=7.5cm]{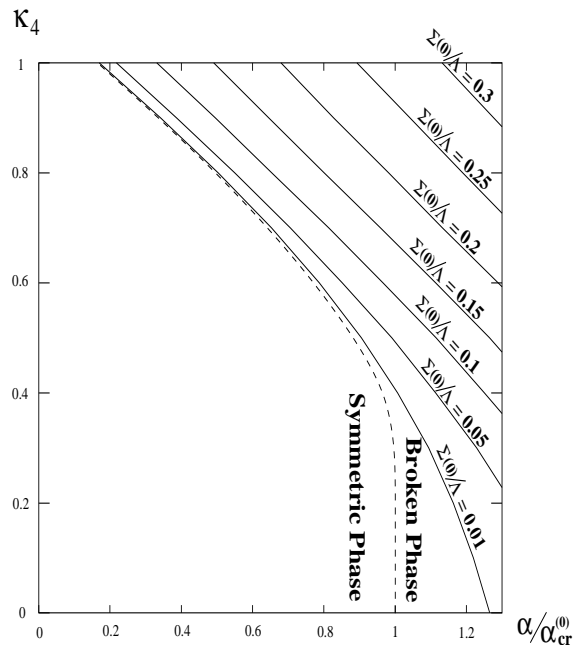}
\end{center}
\caption{Contours of constant $\Sigma(0)/\Lambda$ as a function of 
$\alpha/\alpha^{(0)}_{cr}$ and $\kappa_4$.  Dashed curve is the boundary
between the phases with manifest and spontaneously broken chiral symmetry.} 
\label{Sigma0_contour}
\end{figure}

In Fig. \ref{Sigma0_contour} we display contour curves of equal
$\Sigma(0)/\Lambda$ as a function of $\alpha/\alpha^{(0)}_{cr}$ and $\kappa_4$.
The chiral phase boundary is shown as the dashed curve. For $\kappa_4 < 0.25$,
the critical point $\alpha_{cr}$ for the chiral phase transition is essentially
independent of $\kappa_4$, just as is the case in the gauged NJL model
\cite{kmy,astw,tak4f}. For larger values of $\kappa_4$, this critical point
decreases, and, for $\kappa_4 \simeq 1.1$, this critical point occurs at
$\alpha=0$, i.e., for $\kappa_4 \gsim 1.1$, the local current-current
interaction is sufficient, by itself, to produce a nonzero $\Sigma(0)$, without
any technicolor gauge interaction.  This is again qualitatively similar to the
situation for the gauged NJL interaction.

\section{Expression for $S$ in Terms of Current-Current Correlation Functions}
\label{scorfun}

In this section we review the formulas that we will use to calculate the $S$
parameter in terms of (the derivative of) a certain combination of
current-current correlation functions.  As a measure of corrections to the $Z$
propagator arising from heavy particles and new physics (NP) in theories beyond
the standard model, $S$ was originally defined as \cite{pt}
\beq
S = \frac{4 s_W^2 c_W^2}{\alpha_{em}(m_Z)} \left. \frac{d \Pi^{(NP)}_{ZZ}(q^2)}
{dq^2} \right\vert_{q^2 = 0} \ , 
\label{s}
\eeq
where $s_W^2 = 1-c_W^2 = \sin^2\theta_W$, evaluated at $m_Z$.  More recent
analyses of precision electroweak data define $S$ slightly differently,
replacing the derivative at $q^2=0$ by a finite difference (in the
$\overline{MS}$ scheme) \cite{pdg}
\beq
S_{PDG} = \frac{4 s_W^2 c_W^2}{\alpha_{em}(m_Z)} \Bigg [ 
\frac{\Pi^{(NP)}_{ZZ}(m_Z^2) - \Pi^{(NP)}_{ZZ}(0)}{m_Z^2} \Bigg ] \ . 
\label{spdg}
\eeq
The difference between these definitions is small if the heavy fermion mass
$\Sigma$ satisfies $(2\Sigma/m_Z)^2 \gg 1$, as is the case in the technicolor
models considered here.  To make this quantitative, we recall that in the
one-family technicolor model, $m_W^2 = (g^2/4)f_{TC}^2 (N_c+1) = g^2 f_{TC}^2$,
where $g$ is the SU(2)$_L$ coupling and $f_{TC}$ is the technicolor analogue of
the pion decay constant in QCD.  This yields $f_{TC} \simeq 125$ GeV. We next
use a rough scaling relation connecting the dynamical technifermion mass
\beq
\frac{\Sigma}{\Sigma_{QCD}} \simeq \frac{f_{TC}}{f_\pi} \, 
\Big ( \frac{N_c}{N_{TC}} \Big )^{1/2} \ . 
\label{sigfrel}
\eeq
With $f_\pi = 92.4$ MeV, $\Sigma_{QCD} \simeq m_N/N_c$, and $N_{TC}=2$, one
thus has $\Sigma \simeq 520$ GeV, so that $(2\Sigma/m_Z)^2 \simeq 1.3 \times
10^2$.  For our purposes it will be convenient to use the original definition,
Eq.  (\ref{s}).

Suppressing the SU($N_{TC}$) gauge index, one can write the technifermions 
as a vector, $\psi = (\psi_i,...,\psi_{N_f})$. One can then define vector and 
axial-vector currents as
\beqs
   V_\mu^a(x) &=& \bar\psi(x) T^a \gamma_\mu \psi(x) \cr\cr
   A_\mu^a(x) &=& \bar\psi(x) T^a \gamma_\mu \gamma_5\psi(x) \, ,
\label{vacurrents}
\eeqs
where the $N_f \times N_f$ matrices $T^a$ ($a=1,..., N_f^2-1$) are the
generators of $SU(N_f)$ with the standard normalization.
In terms of these currents, the two-point
current-current correlation functions $\Pi_{VV}$ and $\Pi_{AA}$ are defined via
the equations
\beqs 
&& i\int d^4x\ e^{i q \cdot x}\ \langle 0|T(J_\mu^a(x) J_\nu^b(0))|0 \rangle \
\cr\cr
&=& \delta^{ab}\left(\frac{q_\mu q_\nu}{q^2}-g_{\mu\nu}\right)\Pi_{JJ}(q^2) \ ,
\label{pijjdef}
\eeqs
where $J_\mu^a(x) = V_\mu^a(x), A_\mu^a(x)$.  Since SM gauge interactions are
small at the technicolor scale of several hundred GeV, it follows that the
contributions to $S$ of each of the $N_c$ techniquark electroweak doublets
${U^a \choose D^a}$, $a=1,..,N_c$ and from the technilepton electroweak doublet
${N \choose E}$ are essentially equal.  It is therefore convenient to define a
reduced quantity, $\hat S$, that represents the contribution to $S$ from each
such technifermion doublet, viz.,
\beq
\hat S = \frac{S}{N_D}
\label{shat}
\eeq
where $N_D=N_c+1=4$ for the one-family technicolor theory.  Then, in terms of
the current-current correlation functions defined above, $S$, as defined in
Eq. (\ref{s}), is given by
\beq
\hat S = 4 \pi \left. \frac{d}{d q^2} \left[ \Pi_{VV}(q^2) - \Pi_{AA}(q^2)
   \right] \right\vert_{q^2 = 0}\, ,
 \label{scor}
\eeq
The relation (\ref{scor}) is equivalent to the expression in terms of the
integral over the vector and axial-vector spectral functions for the currents
(\ref{vacurrents}) \cite{dmo},
\beq
\hat S = 4 \pi \int_0^\infty  \frac{ds}{s} \bigg [ \rho_V(s)-\rho_A(s) \bigg ]
\ , 
\label{w0}
\eeq
where 
\beq
\rho_J(s) \equiv \frac{Im(\Pi_{JJ}(s))}{\pi s}
\label{rho}
\eeq
for $J=V,A$.

\section{Calculation of $S$ via Bethe-Salpeter Equation} 
\label{bs}

\begin{figure}
  \begin{center}
    \includegraphics[height=1cm,width=8cm]{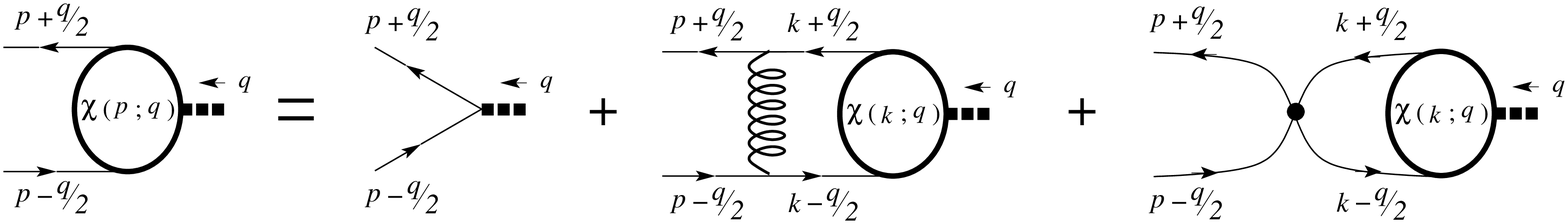}
  \end{center}
\caption{Pictorial representation of the Bethe-Salpeter equation analyzed here.
   The first graph on the right is the bare current vertex, the second
   represents technigluon exchange, and the third represents the contribution
   of the effective local four-fermion operator resulting from the exchange of
   the massive ETC gauge boson $V_{d3}$. See text for further explanation.}
\label{IBSeq}
\end{figure}

The method that we use to calculate $S$ is similar to our earlier work
\cite{hky_s,s}, except that now we use a scattering kernel for the
Bethe-Salpeter equation that includes the exchange of not just the massless
technigluons, but also the massive ETC vector boson $V_{d3}$.  We define
certain Bethe-Salpeter amplitudes $\chi^{(J)}_{\alpha \beta}(p;q,\epsilon)$ as
\beqs 
& & \delta_j^k \left( T^a \right)_{f}^{f'} \int \frac{d^4 p}{(2 \pi)^4} 
\, e^{- i p \cdot r} \, \chi^{(J)}_{\alpha \beta}(p;q,\epsilon) = \cr\cr
& & = \epsilon^\mu \int d^4 x \, e^{i q \cdot x} 
\langle 0 \vert T(\psi^k_{\alpha f}(r/2) \ \bar\psi_{jf'\beta}(-r/2)\
J_\mu^a(x)) | 0 \rangle \ , \cr\cr
& & 
\label{eq:three-point}
\eeqs
where $J=V$ or $A$, and $(f,f')$, $(j,k)$ and $(\alpha, \beta)$ are,
respectively, the flavor, gauge, and spinor indices.  In Fig. \ref{IBSeq} we
show symbolically the terms contributing to the Bethe-Salpeter equation.  
Full propagators are used on the internal technifermion lines and the running
gauge coupling is included.

Closing the fermion legs of the above three-point vertex function and taking
the limit $r \rightarrow 0$, we can express the current-current correlation
function in terms of these amplitudes as
\beqs
& &  \Pi_{JJ}(q^2) = \cr\cr
& & \frac{1}{3} \Bigg ( \frac{N}{2} \Bigg ) \sum_{\epsilon} 
 \int \frac{d^4 p}{i (2\pi)^4} 
{\rm Tr} \left[ \left(\epsilon \cdot G^{(J)}\right) 
   \chi^{(J)}(p;q,\epsilon) \right] \ , \cr\cr
& & 
 \label{eq:Pi_JJ}
\eeqs
where
\beq
G_\mu^{(V)} = \gamma_\mu \ ,\ \quad G_\mu^{(A)} = \gamma_\mu \gamma_5 \ , 
\eeq
and an average has been taken over the polarizations, so that $\Pi_{JJ}(q^2)$
does not depend on the polarization $\epsilon$. We then calculate these
amplitudes and evaluate the requisite combination to obtain $\hat S$.  Since we
include a running coupling in the calculation, we are again working in the
improved ladder approximation.

\begin{figure}
\begin{center}
\includegraphics[height=7cm,width=8.5cm]{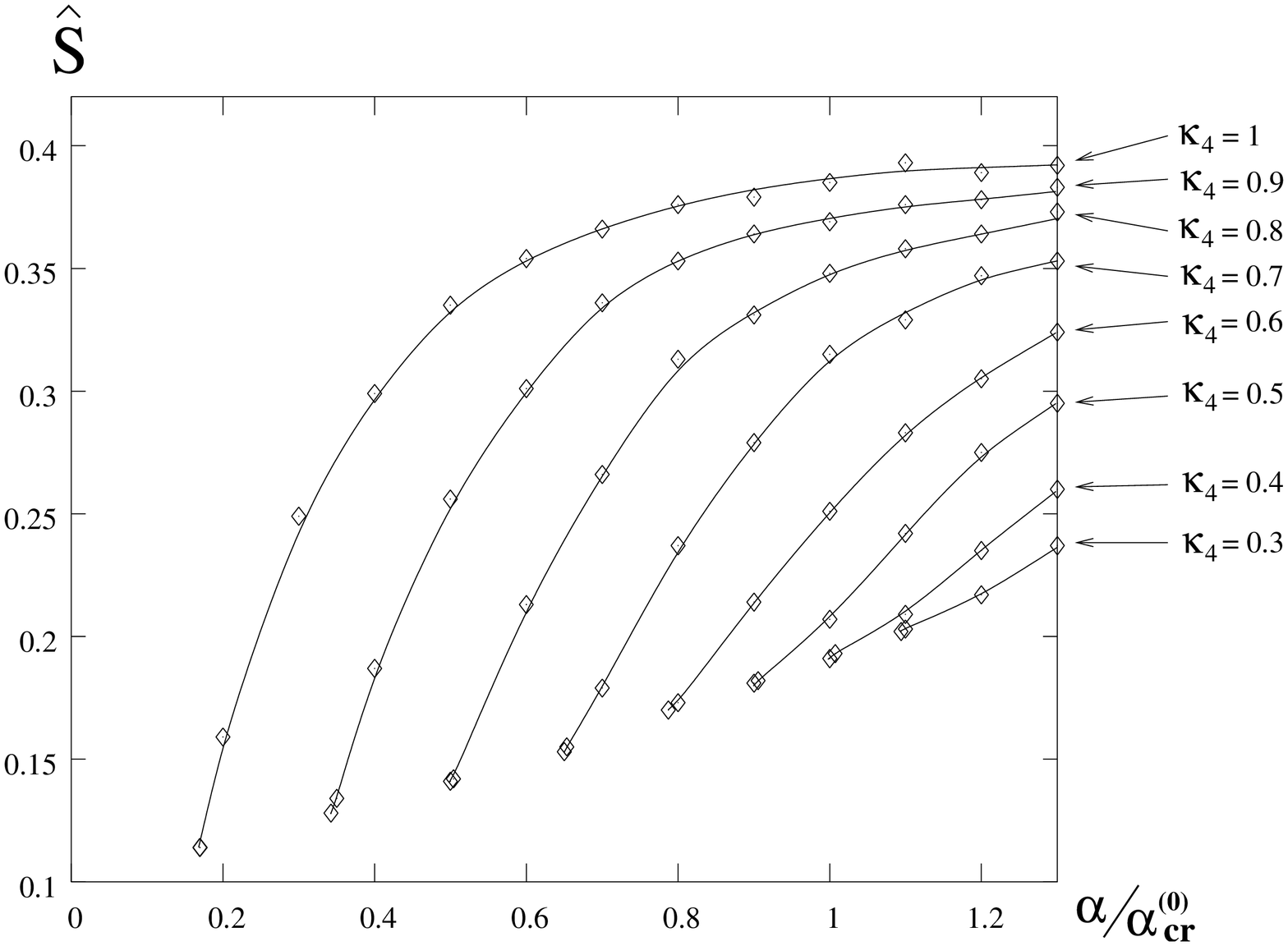}
\end{center}
\caption{$\hat S$ as a function of $\alpha/\alpha^{(0)}_{cr}$ for 
various values of $\kappa_4$.}
\label{S_lambda-fixed} 
\end{figure}

In Fig. \ref{S_lambda-fixed} we show our results for $\hat S$ as a function of
$\alpha/\alpha^{(0)}_{cr}$ for various values of $\kappa_4$.  As in the
analysis of the Schwinger-Dyson equation, here again one formally varies
$\alpha$ and $\kappa_4$ independently, but understands that in a given ETC
theory, they are related, via eqs. (\ref{lambdabar}) and (\ref{nu3}). For a
given value of $\kappa_4$, the smallest value of $\alpha/\alpha^{(0)}_{cr}$
is close to $\alpha_{cr}/\alpha^{(0)}_{cr}$; i.e., the curve starts near to the
chiral boundary.  For fixed $\kappa_4$, $\hat S$ increases as a function of
$\alpha/\alpha^{(0)}_{cr}$.  This is the same trend that we found in
Refs. \cite{hky_s,s} for the pure gauge theory without four-fermion
interaction.  For $\kappa_4=0$, the reason is clear; as $\alpha = \alpha_*$
increases, corresponding to a decrease in $N_f$, one is moving away from the
walking regime toward the more QCD-like regime, and the reduction in $\hat S$
associated with walking behavior is removed.  This increase actually becomes
more abrupt as one increases $\kappa_4$, which indicates that the departure
from walking behavior is more abrupt as one moves away from the chiral phase
boundary in the presence of a substantial four-fermion coupling.  This is in
agreement with what we found for the dynamical technifermion mass, namely that
as one moves into the chirally broken phase from the chiral phase boundary, the
turn-on of $\Sigma$ is more rapid (i.e., the critical exponent $\beta_\Sigma$
is smaller) for larger values of $\kappa_4$, as the exponential suppression
inherent in the form (\ref{sigsol}) is removed.

\begin{figure}
  \begin{center}
    \includegraphics[height=7cm,width=8.5cm]{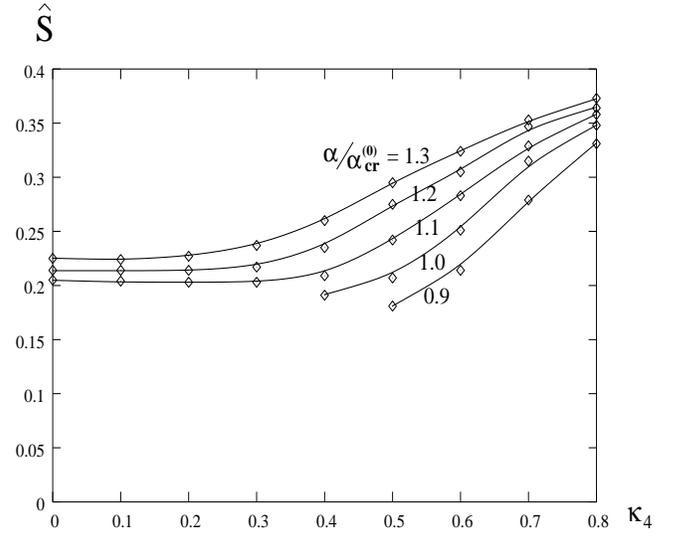}
  \end{center}
\caption{$\hat S$ as a function of $\kappa_4$ for various values of 
$\alpha/\alpha^{(0)}_{cr}$.}
\label{S_alpha-fixed}
\end{figure}
\begin{figure}
  \begin{center}
    \includegraphics[height=9cm,width=8.5cm]{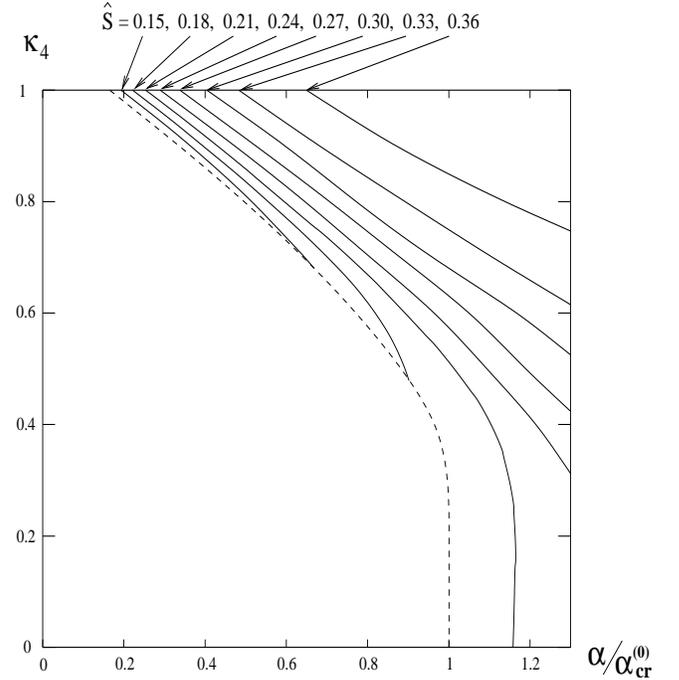}
  \end{center}
\caption{Contours of constant $\hat S$, plotted as functions of 
$\kappa_4$ and $\alpha/\alpha^{(0)}_{cr}$.}
\label{S_contour}
\end{figure}

In Fig. \ref{S_alpha-fixed} we show our results for $\hat S$ as a function of
$\kappa_4$ for a range of values of $\alpha/\alpha^{(0)}_{cr}$.  In a walking
technicolor theory a typical value of $\alpha/\alpha^{(0)}_{cr}$ with $\alpha
\simeq \alpha_*$ could be about 1.1.  In both Fig. \ref{S_lambda-fixed} and
Fig.  \ref{S_alpha-fixed}, the observed feature that increasing $\kappa_4$ at
fixed $\alpha$ eventually increases $\hat S$ is understandable because the
current-current interaction facilitates the formation of the technifermion
condensate and associated appearance of the dynamical mass $\Sigma$.  This is
the same trend as the increase in $\hat S$ that results from increasing
$\alpha_*$ (e.g., by decreasing the number of technifermions, $N_f$), moving
one from the walking regime in the direction of more QCD-like behavior.  From
inspection of the curve for this value, it is evident that $\hat S \simeq 0.2$
for $0 \le \kappa_4 \lsim 0.25$, with $\hat S$ increasing for larger values of
$\kappa_4$.  Since $\kappa_4$ is expected to be rather small
(cf. eq. (\ref{lambdabar_value})), we thus find that in the type of ETC model
considered here, the inclusion of the ETC-induced four-fermion operator has
little effect on $\hat S$ \cite{larger}.  In Fig. \ref{S_contour} we plot
curves of constant $\hat S$ as functions of $\kappa_4$ and
$\alpha/\alpha^{(0)}_{cr}$.  The chiral boundary is again represented by the
dashed curve.

\begin{figure}
  \begin{center}
    \includegraphics[height=7cm]{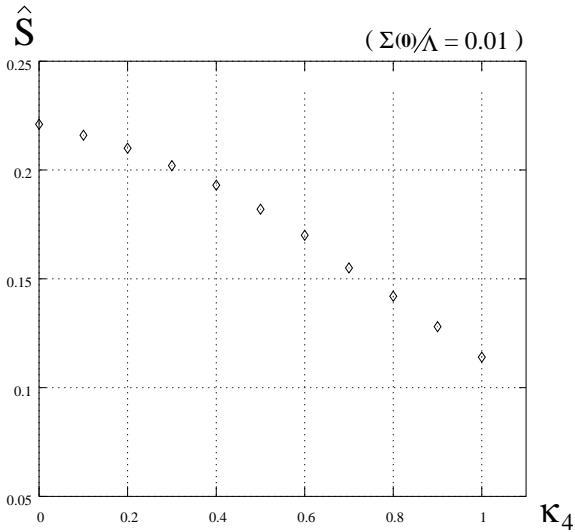}
  \end{center}
\caption{$\hat S$ as a function $\kappa_4$ with $\alpha$ varying so as to
maintain a fixed value of $\Sigma(0)/\Lambda=0.01$.}
\label{S_SigovLamfixed} 
\end{figure}

It is also of interest to investigate how $\hat S$ behaves if one varies both
$\alpha/\alpha^{(0)}_{cr}$ and $\kappa_4$ (formally taken to be an independent
couplings) in such a manner as to move along a contour of a fixed value of the
ratio of physical scales $\Sigma(0)/\Lambda$.  From Fig. \ref{Sigma0_contour},
it follows that the condition of maintaining fixed $\Sigma(0)/\Lambda$ means
that if one increases $\kappa_4$, then this should compensated by a decrease in
$\alpha$.  In models such as those of Ref. \cite{ckm}-\cite{kt}, where walking
behavior extends up to the lowest ETC breaking scale, $\Lambda_3$, the ratio
$\Sigma(0)/\Lambda$ is typically of order 0.1.  The requirement to reduce $S$
as much as possible provides motivation to consider a stronger degree of
walking and hence a smaller value of $\Sigma(0)/\Lambda$.  In Fig.
\ref{S_SigovLamfixed} we present a plot of $\hat S$ for the case where one
keeps this ratio fixed at the value $\Sigma(0)/\Lambda=0.01$.  From inspection
of Fig.  \ref{Sigma0_contour}, one can see that the contour with
$\Sigma(0)/\Lambda=0.01$ intersects the vertical axis at about
$\alpha/\alpha^{(0)}_{cr} \simeq 1.27$ for $\kappa_4=0$, but moves very close
to the chiral boundary as $\kappa_4$ increases past 0.4.  This is in agreement
with the fact that the turn-on of the chiral symmetry-breaking order parameter
$\Sigma$ is more abrupt as $\kappa_4$ increases and the fact that $\hat S$
vanishes in the chirally symmetric phase.  Our results in Fig.
\ref{S_SigovLamfixed} show that $\hat S$ decreases as $\kappa_4$ increases,
with $\alpha/\alpha^{(0)}_{cr}$ reduced so as to keep $\Sigma(0)/\Lambda$
constant.  A plausible interpretation of this is that to remain on the contour
of fixed $\Sigma(0)/\Lambda=0.01$ as $\kappa_4$ increases, one is moving closer
to the chiral boundary.  It is also plausible that this decrease in $\hat S$ is
associated with an increase in the anomalous dimension $\gamma_{\bar\psi\psi}$
for the technifermion bilinear $\bar\psi\psi$.  We recall that
$\gamma_{\bar\psi\psi} = 1$ in the walking limit of a technicolor gauge theory
(and $\gamma_{\bar\psi\psi} \to 2$ for $\alpha=0$, $\kappa_4 \to 1$).  For the
expected small value of $\kappa_4$, our Bethe-Salpeter calculation yields $\hat
S \simeq 0.2$ on this contour, so that $S \simeq 0.8$ (using $N_D=4$ as in
eq. (\ref{nd})).  We recall again the possibility that the dynamics of the
technicolor theory itself could produce large four-fermion operator effects 
which could reduce $S$; here we focus on the ETC-induced contributions.
The above value of $S$ is too large to agree with experimental limits on
$S$, so to maintain the viability of this type of model, it is necessary to
assume that there is a further reduction in $S$.  We comment on this next.

Although one cannot use perturbation theory reliably to calculate $S$ in a
strongly coupled gauge theory, the perturbative formula is often employed for
comparisons of different technicolor models.  The one-loop perturbative
calculation with degenerate fermions having effective masses satisfying
$(2\Sigma/m_Z)^2 \gg 1$ yields the well-known result $S_{pert.}=N_{D,tot.}/(6
\pi)$ where here $N_{D,tot.}=N_{TC} N_D$, so that 
\beq
\hat S_{pert.} = \frac{N_{TC}}{6\pi} \ . 
\label{s_hat_pert}
\eeq
In QCD with just light quarks, and hence $N_D=N_f/2=1$ and hence
$N_{D,tot,QCD}=N_cN_D=3$, this perturbative calculation would predict
$S_{QCD,pert.} \simeq 1/(2\pi) \simeq 0.16$.  To the extent that the
experimental value of $S$ in QCD is dominated by the contributions of
light-quark hadrons in eq. (\ref{w0}) \cite{ud}, it follows that $N_D=1$, so
that $S \simeq \hat S$ for QCD.  This experimental value of $S$ in QCD is $S =
0.33 \pm 0.04$ \cite{svalue}, so that the perturbative estimate is about a
factor of two smaller than the actual value.  An approximate calculation of
$\hat S$ was carried out using the ladder approximation to the Schwinger-Dyson
and Bethe-Salpeter equations for QCD ($N=3$) with $N_f=2$ quarks of negligible
mass \cite{hy_s}.  Studies have also been done for the case where one neglects
the strange quark mass $m_s$, i.e., $N=3$, $N_f=3$ \cite{hy_s,pms}.  Since for
either of these values of $N_f$ the two-loop beta function of the QCD theory
does not exhibit an infrared fixed point, it was necessary in these calculatons
to cut off the growth of the strong coupling.  For typical cutoffs, it was
found that the calculations tended to yield too large a value of $\hat S = S$,
namely $\hat S \simeq 0.45 - 0.5$ \cite{hy_s,pms}.  This suggests that this
type of Bethe-Salpeter calculation may overestimate $S$.  Our studies of $\hat
S$ showed that in a walking theory, $\hat S$ is reduced, relative to its value
in a QCD-like theory \cite{hky_s,s}, in agreement with other studies of the
effect of walking\cite{at_s}-\cite{iwt_s}.  Combining this reduction with a
plausible correction factor to compensate for the tendency of the
Bethe-Salpeter calculation to overestimate $\hat S$ in QCD would yield a
further reduction in $\hat S$.  However, even with a reduction to $\hat S
\simeq 0.1$, inserting the factor $N_D=4$ yields $S \simeq 0.4$, which is
sufficiently large to be of strong concern.  In this context, it should be
noted that the question of the value of $\hat S$ in walking technicolor has
been investigated in recent analyses using holographic methods
\cite{adscft,csaki}, and several authors have found evidence for a a sizable
reduction \cite{adscft}, although Ref. \cite{csaki} did not. An important task
that merits further study is to relate these holographic methods to the sort of
Schwinger-Dyson and Bethe-Salpeter methods used in previous works for pure
walking gauge theories and here for a walking gauge theory with additional
ETC-induced four-fermion interaction.

\section{Summary}
\label{sec:Summary}

Technicolor and extended technicolor theories are very ambitious, since they
aim to explain not only electroweak symmetry breaking and the associated masses
for the $W$ and $Z$ bosons, but also the spectrum of quark, charged lepton, and
neutrino masses.  A successful model of this type would thus explain
longstanding puzzles such as the value of the intergenerational lepton mass
ratio $m_e/m_\mu$ and intragenerational mass ratios such as $m_u/m_d$ and
$m_t/m_b$.  It is thus not surprising that no fully realistic ETC model has
been constructed.  However, since TC/ETC theories continue to provide an
interesting theoretical framework complementary to the standard model itself
and to other approaches such as supersymmetry, it is worthwhile to investigate
their properties further. Accordingly, in this paper, using approximate
numerical solutions of the relevant Schwinger-Dyson and Bethe-Salpeter
equations, we have calculated the correction to the $Z$ boson propagator, as
described by the $S$ parameter, in a technicolor theory, taking account of both
massless technigluon exchange and the dominant contribution from massive ETC
gauge boson exchange. Our results suggest that for the types of ETC models
considered here, this additional contribution from massive ETC gauge boson
exchange has a relatively small effect on $S$.

\bigskip

The research of M. K. and R. S. was partially supported by the grant
NSF-PHY-03-54776.  The research of K. Y. was partially supported by the JSPS
Grant-in-Aid for Scientific Research (B) 18340059, and the Mitsubishi and Daiko
Foundations. We thank Profs. T. Appelquist and T. Takeuchi for helpful 
comments.

\section{Appendix} 

It is useful to compare and contrast the ETC-induced four-fermion operator
whose effects we have analyzed here with the four-fermion operator studied in
the work of Nambu and Jona Lasinio \cite{njl} and to gauged NJL models.  The
NJL model for a single fermion $\psi$ is described by the lagrangian ${\cal
L}_{NJL} = \bar\psi i \fsl{\partial} \psi + {\cal L}_{int,NJL}$, where the
interaction term ${\cal L}_{int,NJL}$ is given by
\beqs
{\cal L}_{int,NJL} & = & 
\frac{c}{2\Lambda^2} \bigg [ 
[\bar\psi \psi ]^2 - [\bar\psi \gamma_5 \psi ]^2 \bigg ] \cr\cr
& = & \frac{2c}{\Lambda^2}[\bar\psi_L \psi_R ][\bar\psi_R \psi_L ] \ , 
\label{njlsp}
\eeqs
Here, $\Lambda$ is a mass scale introduced to make the coupling 
$c$ dimensionless. By a Fierz transformation, this is equivalent to 
\beqs
{\cal L}_{int,NJL} & = & -\frac{c}{4\Lambda^2} \bigg [ 
[\bar\psi \gamma_\mu \psi ][\bar\psi \gamma^\mu \psi ] - 
[\bar\psi \gamma_\mu \gamma_5 \psi ][\bar\psi \gamma^\mu \gamma_5 \psi ] 
\bigg ] 
\cr\cr
& & = -\frac{c}{\Lambda^2}[\bar\psi_L\gamma_\mu \psi_L] 
[\bar\psi_R\gamma^\mu \psi_R] \ . 
\label{njlva}
\eeqs
We define 
\beq
\bar c \equiv \frac{c}{4\pi^2} \ . 
\label{cbar}
\eeq
(This coupling $\bar c$ is identical to the coupling $G\Lambda^2/(4\pi^2)$ in
Ref. \cite{kmy} and to $\beta=G\Lambda^2/(2\pi^2)$ in Ref. \cite{astw}.)  This
theory is invariant under a global chiral symmetry group ${\rm U}(1)_L \times
{\rm U}(1)_R$.  An analysis of the Schwinger-Dyson equation for the fermion,
with an ultraviolet cutoff of $\Lambda$ imposed on the momentum integration,
shows that for $\bar c > 1$, this equation has a nonzero solution for $\Sigma$,
signifying dynamical chiral symmetry breaking.  Associated with this is the
formation of a bilinear fermion condensate $\langle \bar \psi \psi \rangle$
forms, breaking ${\rm U}(1)_L \times {\rm U}(1)_R$ to the diagonal subgroup
${\rm U}(1)_V$.  It is straightforward to generalize the model to an
$N$-component fermion.  

   There has also been interest in considering a class of models in which
$\psi$ transforms as a nonsinglet under an abelian U(1) or non-abelian SU($N$)
gauge group, so that the kinetic term $\bar\psi i \fsl{\partial} \psi$ is
replaced by $\bar\psi i \fsl{D} \psi$, where $D_\mu$ is the appropriate
covariant derivative.  In the abelian case, the Fierz transformation relating
the interaction in eq. (\ref{njlsp}) to that in eq. (\ref{njlva}) applies in
the same manner as for the original NJL model.  In the non-abelian case, the
situation is more complicated.  If the currents in the current-current produce
involve the full non-abelian SU($N$) gauge generators, then a Fierz
transformation operates not just on the Dirac matrices but also on the matrices
representing the gauge generators.  As noted in the text, a number of studies
of a gauged NJL model, particularly in the abelian case, were performed
and mapped out the chiral phase boundary for this type of model as a function
of the gauge coupling $g$ and the NJL coupling $c$ (generically taken to be
independent).

There are both similarities and differences between an ETC-based model of the
type considered in our text and the non-abelian gauged NJL model.  First, 
the ETC interaction yields a product of two vectorial currents, denoted
symbolically as $VV$, rather than the structure of the NJL interaction in 
eq. (\ref{njlva}), which is $VV - AA$.  Second, while the NJL model requires
an ultraviolet cutoff, this is not necessary in the present case since we 
actually start with a reasonably ultraviolet-complete theory, from which the
interaction (\ref{etcjj}) arises as part of the effective low-energy field
theory.  Third, while the general gauged NJL model treats $c$ and the gauge
coupling as independent, this is not the case in a TC/ETC theory, since the TC
gauge group arises as a subgroup of the ETC gauge group and hence the running
TC gauge coupling at a given scale is determined by the ETC gauge coupling at
the higher ETC scales where the ETC gauge degrees of freedom are still active.

In an approximation in which the $\Sigma$ term in the denominator of the
fermion propagator is neglected, thereby linearizing the Schwinger-Dyson
equation, it was found that, as one increases the NJL coupling $\bar c$ from
zero, the critical gauge coupling $\alpha_{cr}$ remains unchanged for $0 \le
\bar c \le 1/4$ and, for $\bar c > 1/4$, the chiral phase boundary can be
described by the following functional relation for the coordinates of a point
$(\alpha_{cr},\bar c_{cr})$ on this boundary,
\beq
\bar c_{cr}=\frac{1}{4}
\bigg (1 + \sqrt{1- \frac{\alpha_{cr}}{\alpha^{(0)}_{cr}}} \ \bigg )^2  \ , 
\label{ceq}
\eeq
or equivalently, $\alpha_{cr}/\alpha^{(0)}_{cr}=4\sqrt{\bar c_{cr}} \,
(1-\sqrt{\bar c_{cr}} \,)$.  As one moves up along the phase boundary from the
point $(\alpha,\bar c)=(\alpha^{(0)}_{cr},0)$ to $(0,1)$, the nature of the
critical singularity also changes from the essential zero in eq. (\ref{sigsol})
to the algebraic singularity (\ref{betadef}) $\Sigma/\Lambda \sim (\bar
c-1)^{1/2}$ as $\bar c \to 1^+$.

In the gauged NJL model, the Schwinger-Dyson equation is, after Euclidean
rotation and angular integration, in the notation of Section \ref{sd},
\beqs
\Sigma(x) & = & \frac{\alpha}{4\alpha^{(0)}_{cr}} \, 
\int_0^{\Lambda^2} y \, dy \frac{\Sigma(y)}{{\rm max}(x,y) \, [y+\Sigma(y)^2]}
\cr\cr
& + &  \frac{\bar c}{\Lambda^2} \int_0^{\Lambda^2} y \, dy \ \frac{\Sigma(y)}
{[y+ \Sigma(y)^2]} 
\label{sdgaugednjl}
\eeqs
If one were formally to replace the factor $\kappa_4(x,y) \, (x+y)/{\rm
max}(x,y)$ in our eqs. (\ref{sigeq})-(\ref{fullkappa}) by $\bar c$, then the
Schwinger-Dyson equation for the ETC model analyzed in the present paper would
be transformed into a structure equivalent to that studied in Refs.
\cite{kmy,astw,tak4f}.  Since the factor $(x+y)/{\rm max}(x,y)$ takes the value
unity for $y \to 0$ and $y \to \infty$ and takes the maximal value 2 (at
$x=y$), one could anticipate that the solutions for the Schwinger-Dyson
equation in the present paper should be formally similar to the results
obtained in Refs. \cite{kmy,astw,tak4f} for the same input values of $\alpha$
and similar values of $\kappa_4$ and $\bar c$.


\begin{thebibliography}{99}

\bibitem{tc}
S. Weinberg, Phys. Rev. D {\bf 19}, 1277 (1979);
L. Susskind, {\it ibid.} D {\bf 20}, 2619 (1979); see also 
S. Weinberg, Phys. Rev. D {\bf 13}, 974 (1976).

\bibitem{etc}
S. Dimopoulos and L. Susskind, Nucl. Phys. B {\bf 155}, 237 (1979);
E. Eichten and K. Lane, Phys. Lett. B {\bf 90}, 125 (1980). 

\bibitem{at94}
T. Appelquist and J. Terning, Phys. Rev. D {\bf 50}, 2116 (1994).

\bibitem{nt}
T. Appelquist and R. Shrock, Phys. Lett. B {\bf 548}, 204 (2002); 
Phys. Rev. Lett. {\bf 90}, 201801 (2003).

\bibitem{ckm}
T. Appelquist, M. Piai, and R. Shrock, Phys. Rev. D {\bf 69}, 015002 (2004);
Phys. Lett. B {\bf 593}, 175 (2004); {\it ibid.} B {\bf 595}, 442 (2004).

\bibitem{kt}
T. Appelquist, N. Christensen, M. Piai, and R. Shrock, Phys. Rev. D {\bf 70}, 
093010 (2004).

\bibitem{wtc1}
B. Holdom, Phys. Lett. B {\bf 150}, 301 (1985).

\bibitem{wtc2}
K. Yamawaki, M. Bando, and K. Matumoto, Phys. Rev. Lett. {\bf 56}, 1335 (1986).

\bibitem{chipt1}
T. Appelquist, D. Karabali, and L. C. R. Wijewardhana, Phys. Rev. Lett. {\bf
57}, 957 (1986).

\bibitem{chipt1a}
T. Appelquist and L. C. R. Wijewardhana, Phys. Rev. D
{\bf 35}, 774 (1987); Phys. Rev. D {\bf 36}, 568 (1987).

\bibitem{bando87}
M. Bando, T. Morozumi, H. So, and K. Yamawaki, Phys. Rev. Lett. {\bf 59}, 389
(1987).

\bibitem{alm}
T. Appelquist, K. Lane, and U. Mahanta, Phys. Rev. Lett. {\bf 61}, 1553 
(1988).

\bibitem{cg}
A. Cohen and H. Georgi, Nucl. Phys. B {\bf 314}, 7 (1989). 

\bibitem{chipt2}
T. Appelquist, J. Terning, and L. C. R. Wijewardhana,
Phys. Rev. Lett.  {\bf 77}, 1214 (1996).

\bibitem{my}
V. Miransky and K. Yamawaki, Phys. Rev. D {\bf 55}, 5051 (1997); {\it ibid.}
{\bf 56}, E 3768 (1997).

\bibitem{sc97}
R. S. Chivukula, Phys. Rev. D {\bf 55}, 5238 (1997).

\bibitem{inst}
T. Appelquist and S. Selipsky, Phys. Lett. B {\bf 400}, 364 (1997). 

\bibitem{shuryak}
M. Velkovsky and E. Shuryak, Phys. Lett. B {\bf 437}, 398 (1998). 

\bibitem{chipt3}
T. Appelquist, A. Ratnaweera, J. Terning, and
L. C. R. Wijewardhana, Phys. Rev. D {\bf 58}, 105017 (1998).

\bibitem{pt}
M. Peskin and T. Takeuchi, Phys. Rev. Lett.  {\bf 65}, 964 (1990);
M. Peskin and T. Takeuchi, Phys.  Rev. D {\bf 46}, 381 (1992). 

\bibitem{ab}
G. Altarelli and R. Barbieri, Phys. Lett. B {\bf 253}, 161 (1991);
G. Altarelli, R. Barbieri, F. Caravaglios,
Int. J. Mod. Phys. A {\bf 13}, 1031 (1998).

\bibitem{pdg}
See http://pdg.lbl.gov and http://lepewwg.web.cern.ch. 

\bibitem{ssvz}
P. Sikivie, L. Susskind, M. Voloshin and V. Zakharov,
Nucl. Phys. B {\bf 173}, 189 (1980).

\bibitem{scalc}
B. Holdom and J. Terning, Phys. Lett. B {\bf 247}, 88 (1990);
M. Golden and L. Randall, Nucl. Phys. {\bf B} {\bf 361}, 3 (1991);
H. Georgi, Nucl. Phys. B {\bf 363}, 301 (1991);
R. Johnson, B.-L. Young, and D. McKay, Phys. Rev. D {\bf 43}, R17 (1991);
R. Cahn and M. Suzuki, Phys. Rev. D {\bf 44}, 3641 (1991);

\bibitem{at_s}
T. Appelquist and G. Triantaphyllou, Phys. Lett. B {\bf 278}, 345 (1992).

\bibitem{acd_s}
R. Sundrum and S. Hsu, Nucl. Phys. B {\bf 391}, 127 (1993).

\bibitem{hy_s}
M.~Harada and Y.~Yoshida,
Phys. Rev. D {\bf 50}, 6902 (1994). 

\bibitem{aes96}
T. Appelquist, N. Evans, and S. Selipsky, Phys. Lett. B {\bf 374}, 145 (1996).

\bibitem{as_s}
T. Appelquist and F. Sannino, Phys. Rev. D {\bf 59}, 067702 (1999).

\bibitem{iwt_s}
S. Ignjatovic, L. C. R. Wijewardhana, and T. Takeuchi, Phys. Rev. 
D {\bf 61}, 056006 (2000).

\bibitem{sml}
N. Christensen and R. Shrock, Phys. Rev. Lett. {\bf 94}, 241801 (2005).

\bibitem{hky_s}
M. Harada, M. Kurachi, and K. Yamawaki, Prog. Theor. Phys. {\bf 115}, 765
(2006); see also 
M. Kurachi, in {\it Proc. of the 2004 Workshop on Dynamical Symmetry Breaking,
  Dec. 2004}, eds. M. Harada and K. Yamawaki (Physics Department, Nagoya 
University, Nagoya), p. 125.

\bibitem{s}
M.~Kurachi and R.~Shrock,
Phys. Rev. D {\bf 74}, 056003 (2006).

\bibitem{sg}
M.~Kurachi and R.~Shrock,
JHEP {\bf 12}, 034 (2006).

\bibitem{adscft}
D. K. Hong and H.-U. Yee, Phys. Rev. D {\bf 74}, 015011 (2006);
J. Hirn and V. Sanz, Phys. Rev. Lett. {\bf 97}, 121803 (2006);
J. Hirn and V. Sanz, JHEP {\bf 0703}, 100 (2007); M. Piai, hep-ph/0608241.

\bibitem{csaki}
K. Agashe, C. Cs\'aki, C. Grojean, and M. Reece, arXiv:0704.1821. 

\bibitem{njl}
Y. Nambu and G. Jona-Lasinio, Phys. Rev. {\bf 122}, 345 (1961); 
{\it ibid.} {\bf 124}, 246 (1961). 

\bibitem{jbw}
K. Johnson, M. Baker, and R. Willey, Phys. Rev. {\bf 136}, B1111 (1964);
M. Baker and K. Johnson, {\it ibid.} D {\bf 3}, 2516 (1971). 

\bibitem{gn}
D. Gross and A. Neveu, Phys. Rev. D {\bf 10}, 3235 (1974).

\bibitem{hooft}
G. 't Hooft, Phys. Rev. Lett. {\bf 37}, 8 (1976); 
Phys. Rev. D {\bf 14}, 3432 (1976). 

\bibitem{qcd_instanton}
D. Caldi, Phys. Rev. Lett. {\bf 39}, 121 (1977);
C. Callan, R. Dashen, D. Gross, Phys. Rev. D {\bf 17}, 2717 (1978);
R. Carlitz, Phys. Rev. D {\bf 17}, 3225 (1978); R. Carlitz and Creamer,
Ann. Phys. (N.Y.) {\bf 118}, 429 (1979).  

\bibitem{qed4f}
W. Bardeen, C. Leung, and S. Love, Phys. Rev. Lett. {\bf 56}, 1230 (1986);
C. Leung, S. Love, and W. Bardeen, Nucl. Phys. B {\bf 273}, 649 (1986);
C. Leung, S. Love, and W. Bardeen, Nucl. Phys. B {\bf 323}, 493 (1989). 

\bibitem{fs}
E. Farhi and L. Susskind, Phys. Rept. {\bf 74}, 277 (1981). 

\bibitem{nambu88}
Y. Nambu, in {\it Proc. XI Warsaw Symposium on Elementary Particle Physics},
ed. Z. Ajduk et al., (World Scientific, Singapore, 1989), p. 1; 
Y. Nambu, in {\it 1988 International Workshop on Strongly Coupled Gauge 
Theories}, ed. M. Bando et al. (World Scientific, Singapore, 1989), p. 3. 

\bibitem{kmy}
K.-I. Kondo, H. Mino, and K. Yamawaki, Phys. Rev. D {\bf 39}, 2430 (1989);
K. Yamawaki, in {\it Proc. 12th Johns Hopkins Workshop on Current Problems in
Particle Theory, 1988}. 

\bibitem{astw}
T. Appelquist, M. Soldate, T. Takeuchi, and L. C. R. Wijewardhana, 
in {\it Proc. 12th Johns Hopkins Workshop on Current Problems in
Particle Theory, 1988}. 

\bibitem{my4f}
V. A. Miransky and K. Yamawaki, Mod. Phys. Lett. A {\bf 4}, 129 (1989).

\bibitem{mty89}
V. A. Miransky, M. Tanabashi, and K. Yamawaki, Phys. Lett. B {\bf 221}, 177
(1989). 

\bibitem{aetw}
T. Appelquist, M. Einhorn, T. Takeuchi, and L. C. R. Wijewardhana, 
Phys. Lett. B {\bf 220}, 223 (1989). 

\bibitem{mny}
V. A. Miransky, T. Nonoyama, and K. Yamawaki, Mod. Phys. Lett. A {\bf 4}, 1409
(1989). 

\bibitem{holdom4f}
B. Holdom, Phys. Lett. B {\bf 226}, 137 (1989). 

\bibitem{tak4f}
T. Takeuchi, Phys. Rev. D {\bf 40}, 2697 (1989). 

\bibitem{atew}
T. Appelquist, T. Takeuchi, M. Einhorn, and L. C. R. Wijewardhana,
Phys. Lett. B {\bf 232}, 211 (1989). 

\bibitem{bhl}
W. Bardeen, C. Hill, and M. Lindner, Phys. Rev. D {\bf 41}, 1647 (1990). 

\bibitem{ashapira}
T. Appelquist and O. Shapira, Phys. Lett. B {\bf 249}, 83 (1990). 

\bibitem{amnw}
T. Appelquist, U. Mahanta, D. Nash, and L.C.R.  Wijewardhana,
Phys. Rev. D {\bf 43}, 646 (1991).

\bibitem{ksy}
K. I. Kondo, S. Shuto, and K. Yamawaki, Mod. Phys. Lett. A {\bf 6}, 3385
(1991). 

\bibitem{aoki99}
K. I. Aoki, K. Morikawa, J. I. Sumi, H. Terao, and M. Tomoyose, Prog. Theor. 
Phys. {\bf 102}, 1151 (1999). 

\bibitem{hr} 
Models with technifermions transforming according to higher-dimensional
representations of the technicolor gauge group have been considered, e.g., in
\cite{el88}-\cite{ts}. The simplest case, in which the technifermions are in
the fundamental representation, will suffice for our present study.

\bibitem{el88}
E. Eichten and K. Lane, Phys. Lett. B {\bf 222}, 274 (1988).

\bibitem{sannino}
D. Hong, S. Hsu, and F. Sannino, Phys. Lett. B {\bf 597}, 89 (2004);
F. Sannino and K. Tuominen, Phys. Rev. D {\bf 71}, 051901 (2005).

\bibitem{ts}
N. Christensen and R. Shrock, Phys. Lett. B {\bf 632}, 92 (2006).

\bibitem{integer}
%
Here and below, when we mention non-integral values of $N_f$, it is implicitly
understood that physical values of $N_f$ are, of course, non-negative integers,
and the non-integral values are defined via an analytic continuation away from
these physical values.

\bibitem{lgt}
Y.  Iwasaki et al., Phys. Rev. Lett. {\bf 69}, 21 (1992)
Phys. Rev. D {\bf 69}, 014507 (2004);
P. Damgaard, U. Heller, A. Krasnitz, and P. Olesen, Phys. Lett. B {\bf 400},
169 (1997);
R. Mawhinney, Nucl. Phys. B (Proc. Suppl.) {\bf 83}, 57 (2000).

\bibitem{cjt}
J. Cornwall, R. Jackiw, and E. Tomboulis, Phys. Rev. D {\bf 10}, 2428 (1974);
K. Lane, Phys. Rev. D {\bf 10}, 2605 (1974); 
T. Maskawa and H. Nakajima, Prog. Theor. Phys. {\bf 52}, 1326 (1974); 
{\it ibid.} {\bf 54}, 860 (1975); 
R. Fukuda and T. Kugo, Nucl. Phys. B {\bf 117}, 250 (1974). 

\bibitem{gardi}
E. Gardi and M. Karliner, Nucl. Phys. B {\bf 529}, 383 (1998);
E.~Gardi, G.~Grunberg and M.~Karliner, JHEP {\bf 07}, 007 (1998).

\bibitem{exactsolution}
R.~Corless, G.~Gonnet, D.~Hare, D.~Jeffrey and D.~Knuth,
Adv.\ Comput.\ Math.\  {\bf 5}, 329 (1996).

\bibitem{qcdnjl}
Some reviews include U. Vogl and, W. Weise, Prog. Part. Nucl. Phys. {\bf 27}, 
195 (1991); S. Klevansky, Rev. Mod. Phys. {\bf 64}, 649 (1992); 
V. A. Miransky, {\it Dynamical Symmetry Breaking in Quantum Field Theories} 
(World Scientific, Singapore, 1993). 

\bibitem{kogut}
J. Kogut, E. Dagotto, A. Kocic, Phys. Rev. Lett. {\bf 61}, 2416 (1988); 
Phys. Rev. Lett. {\bf 62}, 1001 (1989); Phys. Rev. D {\bf 43}, R1763 (1991).
S. Hands, J. Kogut, and E. Dagotto, Nucl. Phys. B {\bf 333}, 551 (1990). 

\bibitem{schierholz}
M. G\"ockeler, R. Horsley, E. Laermann, P. Rakow, G. Schierholz, 
R. Sommer, and U. Wiese, Nucl. Phys. B {\bf 334}. 527 (1990); 
M. G\"ockeler, R. Horsley, E. Laermann, U. Wiese, P. Rakow, G. Schierholz,
and R. Sommer, Phys. Lett. B {\bf 251}, 567 (1990); 
Nucl. Phys. B {\bf 371}, 713 (1992). 

\bibitem{mmw}
M.~Harada, M.~Kurachi and K.~Yamawaki,
Phys.\ Rev.\ D {\bf 68}, 076001 (2003); 

\bibitem{pms}
M.~Harada, M.~Kurachi and K.~Yamawaki,
Phys.\ Rev.\ D {\bf 70}, 033009 (2004).

\bibitem{dmo}
T.~Das, V.~S.~Mathur and S.~Okubo,
Phys. Rev. Lett. {\bf 19}, 859 (1967).

\bibitem{larger}
We note again the possibility that the dynamics of the technicolor theory
itself could produce large four-fermion operator effects, which would affect
$S$. Here we focus on the ETC-induced contributions. 

\bibitem{ud}
This is consistent with the fact that the contributions of the light-quark 
vector and axial-vector mesons $\rho$ and $a_1$ largely saturate the 
spectral function integral for $S$, eq. (\ref{w0}), in QCD. 

\bibitem{svalue}
J. Gasser and H. Leutwyler, Nucl. Phys. B {\bf 250}, 465 (1985);
{\it ibid.} B {\bf 250}, 517 (1985); 
M. Harada and K. Yamawaki, Phys. Rept. {\bf 381}, 1 (2004).


\end{thebibliography}
\end{document}